\documentclass[arguments]{aastex631}

\usepackage{hyperref}
\usepackage{amsmath}
\usepackage{enumerate}
\usepackage[utf8]{inputenc}
\usepackage{cleveref}
\usepackage{amssymb}
\usepackage[T1]{fontenc}
\usepackage{enumitem}
\usepackage{url}
\usepackage[misc]{ifsym}
\usepackage{multirow}
\usepackage{mhchem}
\usepackage{booktabs}
\usepackage{subfigure}

\begin{document}

\title{Dual-Backend Multibeam Position Switching Targeted SETI Observations toward Nearby Active Planet-hosting Systems with FAST}

\author[0000-0002-1190-473X]{Jian-Kang Li}
\affiliation{Institute for Frontiers in Astronomy and Astrophysics, Beijing Normal University, Beijing 102206, People's Republic of China}
 \affiliation{School of Physics and Astronomy, Beijing Normal University, Beijing 100875, People's Republic of China; \url{tjzhang@bnu.edu.cn}}
\author[0000-0002-1190-473X]{Zhen-Zhao Tao}
 \affiliation{Institute for Astronomical Science, Dezhou University, Dezhou 253023, People's Republic of China}
 \affiliation{College of Computer and Information, Dezhou University, Dezhou 253023, People's Republic of China}
 \author[0000-0002-3386-7159]{Pei Wang}
 \affiliation{National Astronomical Observatories, Chinese Academy of Sciences, Beijing 100101, People's Republic of China}
\author[0000-0002-3363-9965]{Tong-Jie Zhang\href{mailto:tjzhang@bnu.edu.cn}{\textrm{\Letter}}}
\affiliation{Institute for Frontiers in Astronomy and Astrophysics, Beijing Normal University, Beijing 102206, People's Republic of China}
\affiliation{School of Physics and Astronomy, Beijing Normal University, Beijing 100875, People's Republic of China; \url{tjzhang@bnu.edu.cn}}



\begin{abstract}
The Five-hundred-meter Aperture Spherical Telescope (FAST), the world's largest single-dish radio telescope, lists the search for extraterrestrial intelligence (SETI) as one of its key scientific objectives. In this work, we present a targeted SETI observation for 7 nearby active stars utilizing the FAST L-band multibeam receiver, employing a observational strategy that combines position switching with multibeam tracking to balance on-source integration time with the accuracy of the beam response. Using both pulsar and SETI backends, we perform a comprehensive search for narrowband drifting signals with Doppler drift rates within diversified drift rate ranges and channel-width periodic signal with periods between 0.12 and 100 s and duty cycles between 10\% and 50\%. No credible radio technosignatures were detected from any of the target systems. Based on this null result, we place constraints on the presence of transmitters at a 95\% confidence level, ruling out narrowband transmitters with with EIRP above $3.98\times10^8 \,\mathrm{W}$ and periodic transmitter with EIRP above $1.80\times10^{10} \,\mathrm{W}$,respectively, within the observation band.

\end{abstract}

\keywords{\href{http://astrothesaurus.org/uat/2127}{Search for extraterrestrial intelligence (2127)}; \href{http://astrothesaurus.org/uat/1338}{Radio astronomy (1338)};\href{http://astrothesaurus.org/uat/498}{Exoplanets (498)}\href{http://astrothesaurus.org/uat/74}{Astrobiology (74)}}


\section{Introduction}\label{sec:introduction}
The Search for Extraterrestrial Intelligence (SETI) endeavors to detect evidence of technologically advanced civilizations beyond Earth by looking for technosignatures, especially in radio frequency.  To maximize the chances of discovery, many SETI programs focus on targeted observations of selected targets. In practice, this means pointing sensitive radio telescopes at certain celestial locations where the potential for detecting an artificial signal can be enhanced, such as planet-hosting stars \citep{2013ApJ...767...94S,2016ApJ...827L..22T,2016AJ....152..181H,2017ApJ...849..104E,2019AJ....157..122P,2020AJ....159...86P,2020AJ....160...29S,2021AJ....161..286T,2021NatAs...5.1148S,2021AJ....161...55M,2022ApJ...932...81G,2022AJ....164..160T,2023AJ....166..190T,2025AJ....169..217L}, Galactic center \citep{2020PASA...37...35T,2021AJ....162...33G,2024AJ....168..284B}, and nearby galaxies \citep{2017AJ....153..110G,2024AJ....167...10C}. Despite decades of targeted SETI observations, no confirmed technosignature has been found, underscoring the need for continued searches with ever-improving instruments and strategies.

The detection for thousands of exoplanets over the past three decades reflects a transformative period in astronomy, revealing that planetary systems are ubiquitous around most stars in the Milky Way. \cite{2012Natur.481..167C} suggested that, in our Galaxy, at least one planet per star averagly, and \cite{2013PNAS..11019273P} found that approximately 22\% of Sun-like stars contain an Earth-sized planet within their temperate habitable zones (HZ). Many of the known exoplanets in or near HZ orbit M dwarfs, which are smaller and cooler than the Sun, and consist about 75\% of all stars in the Galaxy. Notably, such stars often exhibit stronger magnetic fields \citep{1996ApJ...459L..95J,2009ApJ...692..538R,2017NatAs...1E.184S,2021A&ARv..29....1K} and higher activity levels \citep{2007AcA....57..149K}. Particularly, they can powerful flares and bursts from X-rays to radio waves \citep{2024LRSP...21....1K}, hence affect the planetary habitability. Strong and frequent flare stellar activities and coronal mass ejections (CMEs) can erode planetary atmospheres, making the planet uninhabitable \citep{2007AsBio...7..167K,2016ApJ...826..195K,2017ApJ...841..124V,2017ApJ...844L..13G}. On the other hand, it is still debated that tidally locked planets might still retain habitable for some certain suitable conditions \citep{2007AsBio...7...30T,2013ApJ...771L..45Y,2023NatCo..14.2125W}. Thus, M dwarf star systems remain central to exoplanet habitability studies and SETI target lists. Beyond M dwarfs, other types of star can also provide habitable environments around their orbit. Closely adjacent to the Sun on the stellar spectrum are K-type and F-type stars, both recognized as potentially promising hosts for life-bearing worlds. K dwarfs, comprising about 15\% of main-sequence stars, are also known as ``Goldilock'' stars since they can exhibit moderate stellar activity, as well as relatively stable and gentle radiation environments, which enhance their suitability for hosting planets with stable atmospheres and climates \citep{1999Icar..141..399C,2016ApJ...827...79C}. F-type stars, though comparatively rare, possess broader and farther HZ allowing Earth-like planets to avoid tidal locking \citep{2017AN....338..413S}, while their stronger ultraviolet radiation could be harmful to DNA \citep{2014IJAsB..13..244S}. Although F-type stars are paid less habitability consideration, the above circumstellar environments features are still potential for life existstence \citep{2024ApJS..274...20P}. These considerations suggest that F, K, and M stars are promising SETI targets, although they have received relatively less observational emphasis compared with G-type dwarfs.

The unprecedented sensitivity of FAST \citep{2006ScChG..49..129N,2011IJMPD..20..989N,2016RaSc...51.1060L,2020RAA....20...64J} provides a valuable opportunity to expand SETI research \citep{2020RAA....20...78L,2021RAA....21..178C}. Previous SETI efforts with FAST have primarily focused on looking for continuous narrow-band signals using the 19-beam L-band receiver in a single-pointing tracking mode. A given star system is observed by one of the beams while the others serve as reference beams and the candidate should only appear in the on-target beam but not in the reference beams \citep{2022AJ....164..160T,2023AJ....166..190T,2023AJ....165..132L,2025AJ....169..217L,2023AJ....166..245H}, which can reject a considerable part of radio frequency interference (RFI). Neverthless, the gains and leakages of different beams can vary remarkably, which may lead ambiguity in distinguish candidate and RFI. These earlier FAST observations employed single dedicated SETI backend for narrowband signal expected to be Doppler-drifted (\citealp{2019ApJ...884...14S}; \citealp{2022ApJ...938....1L}; \citealp{2023AJ....166..182L}). Recently, It has been suggested that broadband signal might be preferred than narrowband signal duw to the terrestrial economics of beacon transmitter \citep{2010AsBio..10..491B,2010AsBio..10..475B}, the employing of frequency-shift keying \citep{2011AcAau..69..777F} and the consideration of robustness to RFI \citep{2012AcAau..78...80M,2012AcAau..81..227M}. In observations, \cite{2021AJ....162...33G,2022ApJ...932...81G} have carried out boardband signal search in SETI observation by identifying artificial dispersion, and \cite{2023AJ....165..255S} have searched the periodic technosignatures by the repeating period of the signal. Overall, the fundamental principle in SETI is to identify signals that are distinguishable from natural astrophysical emission, not merely limited to narrowband features.

In this paper, we propose a combination of multibeam method and position switching observations. The central beam serves as on-target beam and the edge beams serve as reference beams in the on-source observation. When switching telescope direction, one of the edge beam point at the target source to serve as on-target beam, while the other beams serve as reference beam. The instrumental factors can be calibrated by on-off switching, and the RFI rejection efficiency can be enhanced both by on-off and multibeam. Furthermore, We also utilize pulsar (psr) backend and SETI backend to record high temporal resolution data and high-frequency resolution data simultaneously during SETI observations. The dual-backend approach allows us to conduct two complementary searches on the same observation targets. The high-frequency resolution data is for the conventional narrowband, continuous-wave drifting signals, while the high temporal resolution data is for channel-width periodic signal search.   
In section \ref{sec:Observations}, we discuss the observation targets and strategy in this work. The data process pipeline and princeple are introduced in section \ref{sec:DataAnalysis}, and the corresponding results are presented in section \ref{sec:Results}. Section \ref{sec:Discussion} is the implication of our results and section \ref{sec:Conclusion} lists the conclusion of this work.

\section{Observations}\label{sec:Observations}
\subsection{Targets}\label{subsec:Targets}
\begin{figure}[thpb]
  \centering
  \includegraphics[width=0.65\textwidth]{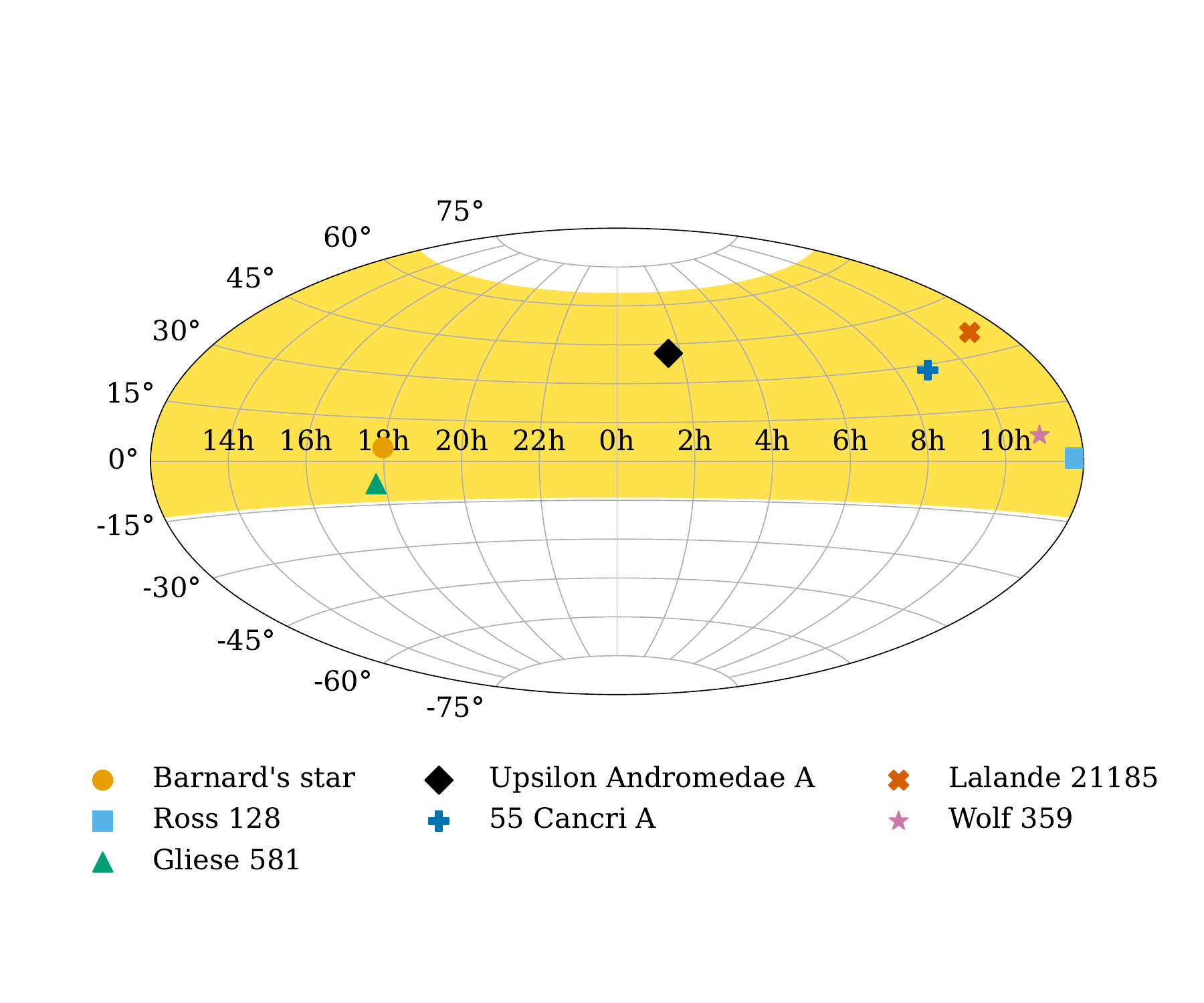}
  \caption{\label{fig:Source_coordinate}  Observation targets in equatorial coordinates. The observable sky coverage of FAST is filled by yellow color.}
\end{figure}

We prioritize 7 stellar systems for our SETI survey based on their astrobiological potential, observability, as well as relative proximities to Earth. Targets are chosen for their high astrobiological potential to increase the likelihood of life. Nearby stellar systems are prioritized to enhance signal detectability. We also confirme that all targets are practically observable, with celestial positions falling within the accessible sky of our telescope. A comprehensive summary of these targets is provided in Table \ref{table:Target_info}, with their sky distribution visualized in Figure \ref{fig:Source_coordinate}. We also observe 3C286, 3C48 and 3C147 as flux calibrators.

\begin{longrotatetable}
\begin{deluxetable}{lccccccc}
\tablecaption{Information of the Observation Targets in This Work.\label{table:Target_info}}
\tablehead{
\colhead{Source Name} & 
\colhead{R.A. (J2000.0)\textsuperscript{(1)}} & 
\colhead{Decl. (J2000.0)\textsuperscript{(1)}} & 
\colhead{Distance (pc)\textsuperscript{(1)}} & 
\colhead{Spectral Type} & 
\colhead{Luminosity ($L_\odot$)} & 
\colhead{Temperature (K)} & 
\colhead{Exoplanet Candidate}
}
\startdata
Barnard's Star         & 17:57:48.50 & +04:41:36.11 & 1.83  & M4V\textsuperscript{(2)}     & 0.00340\textsuperscript{(9)}   & 3195\textsuperscript{(13)}    & Barnard's Star b, c, d, e\textsuperscript{(15)} \\
Ross 128               & 11:47:44.40 & +00:48:16.40 & 3.37  & M4V\textsuperscript{(3)}     & 0.00366\textsuperscript{(9)}   & 3189\textsuperscript{(9)}    & Ross 128 b\textsuperscript{(16)} \\
Gliese 581             & 15:19:26.83 & -07:43:20.19 & 6.30  & M3V\textsuperscript{(4)}     & 0.012365\textsuperscript{(10)}   & 3500\textsuperscript{(10)}    & Gliese 581 b, c , d, e, f, g\textsuperscript{(17)} \\
Upsilon Andromedae A   & 01:36:47.84 & +41:24:19.65 & 13.48 & F8V\textsuperscript{(5)}     & 3.1\textsuperscript{(11)}   & 6614\textsuperscript{(11)}    & Upsilon Andromedae A b, c, d\textsuperscript{(18)}  \\
55 Cancri A            & 08:52:35.81 & +28:19:50.96 & 12.59 & K0IV-V\textsuperscript{(6)}  & 0.617\textsuperscript{(12)}   & 5172\textsuperscript{(14)}    & 55 Cancri A b, c, d, e, f\textsuperscript{(19)}  \\
Lalande 21185          & 11:03:20.19 & +35:58:11.58 & 2.55  & M2V\textsuperscript{(7)}     & 0.02194\textsuperscript{(9)}   & 3547\textsuperscript{(9)}    & Lalande 21185 b, c, d\textsuperscript{(20)}  \\
Wolf 359               & 10:56:28.92 & +07:00:53.00 & 2.41  & M6V\textsuperscript{(8)}     & 0.00106\textsuperscript{(9)}   & 2749\textsuperscript{(9)}    & Wolf 359 b\textsuperscript{(21)}  \\
\enddata
\end{deluxetable}
\tablerefs{
(1) \cite{2023A&A...674A...1G}; 
(2) \cite{1997AJ....113..806G}; 
(3) \cite{2004AAS...205.5503G}; 
(4) \cite{2005A&A...443L..15B}; 
(5) \cite{2009ApJS..180..117A}; 
(6) \cite{2003AJ....126.2048G}; 
(7) \cite{1989ApJS...71..245K}; 
(8) \cite{1994AJ....108.1437H}; 
(9) \cite{2021ApJ...918...40P};
(10) \cite{2024A&A...688A.112V};
(11) \cite{2021AJ....162..198B};
(12) \cite{2024A&A...682A.145S};
(13) \cite{2024A&A...690A..79G};
(14) \cite{2018A&A...619A...1B};
(15) The four planets orbiting Barnard's star are proposed by \cite{2024A&A...690A..79G} and comfirmed by \cite{2025ApJ...982L...1B}; 
(16) Ross 128 b is discovered by \cite{2018A&A...613A..25B}, and \cite{2024A&A...690A.234L} comfirms that this planet retains status as hosting lonely; 
(17) Currently, the existstence of Gliese 581 b, c and e have been comfirmed \citep{2014Sci...345..440R,2018A&A...609A.117T,2024A&A...688A.112V}, the existstence of Gliese 581 f and g are refuted \citep{2014Sci...345..440R,2013ApJ...764....3R}, and the existstence of Gliese 581 d still remains doubtful and under vigorous debate as it is thought to be a false positive result from stellar activity \citep{2014Sci...345..440R,2015MNRAS.452.2745S,2016A&A...585A.144H,2018A&A...609A.117T,2022AJ....163..169D,2024A&A...688A.112V}, whlie some studies still support its existstence \citep{2012AN....333..561V,2016A&A...585A.144H,2024RNAAS...8...20C}; 
(18) Upsilon Andromedae A has once been thought to host four planets \citep{2008IAUS..249..469B,2011A&A...525A..78C}, but the existstence of Upsilon Andromedae e is still suggested to be instrumental artifact \citep{2014ApJ...795...41M,2015ApJ...798...46D}; 
(19) The 55 Cancri A planetary system is confirmed by \cite{1997ApJ...474L.115B,2002ApJ...581.1375M,2004ApJ...614L..81M,2008ApJ...675..790F}, and it is suggested that a hypothetical planet g exist in the gap between planets f and d \cite{2008ApJ...689..478R}; 
(20) Lalande 21185 b is comfirmed by \cite{2020A&A...643A.112S} and Lalande 21185 c is comfirmed by \cite{2021ApJS..255....8R}, planet d is suspected to orbit between planet b and c \cite{2022AJ....163..218H}; 
(21)  Wolf 359 b is reported by \cite{2019arXiv190604644T} while its existence is unable to be either confirmed or refuted \citep{2023AJ....166..260B}.
}
\end{longrotatetable}

\subsubsection{Planetary Habitability}
\begin{figure}[thpb]
  \centering
  \includegraphics[width=0.8\textwidth]{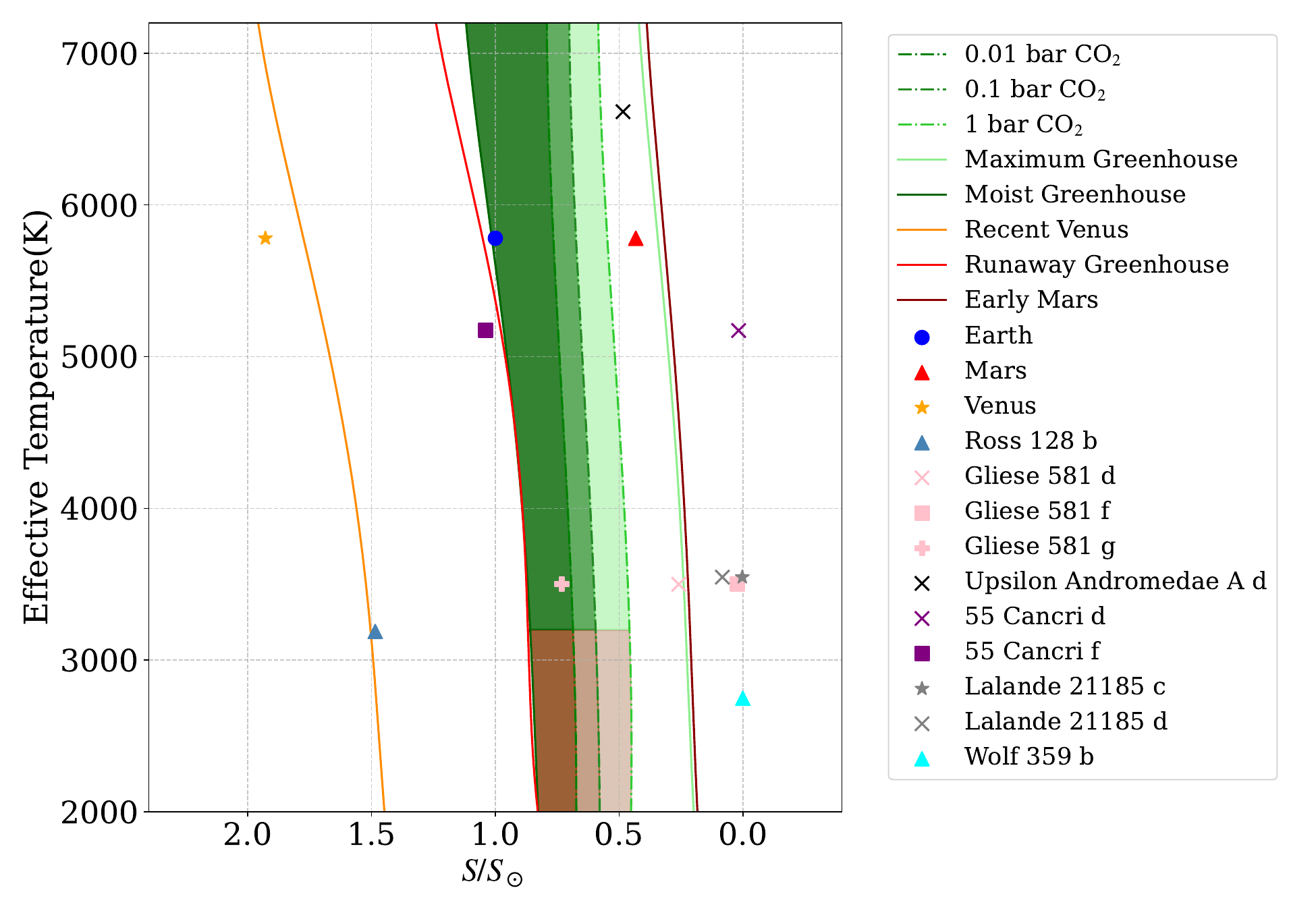}
  \caption{\label{fig:habitablezone}  Various HZ boundaries for stars with different effective temperatures. The five solid lines are the HZ boundaries determined by stellar flux HZ model of \cite{2013ApJ...765..131K}, and the green and brown shaded regions are the HZCL model of \cite{2019ApJ...878...19S} assuming limiting \ce{CO2} concentrations of 0.01 bar (dark green), 0.1 bar (lighter green), and 1 bar (lightest green). The brown contours denote regions around low-temperature stars where photochemical effects could allow CO concentrations to exceed short-term human safety thresholds ($>100$~ppm) at the moist greenhouse boundary, under an assumed surface flux of $3 \times 10^{11}$ molecules~cm$^{-2}$~s$^{-1}$. The planetary parameters are from \cite{2025ApJ...982L...1B} for Barnard's star, \cite{2024A&A...690A.234L} for Ross 128, \cite{2024A&A...688A.112V,2010ApJ...723..954V} for Gliese 581, \cite{2012A&A...545A...5L} for Upsilon Andromedae A, \cite{2010ApJ...722..937D,2011ApJ...737L..18W,2018A&A...619A...1B} for 55 Cancri A, \cite{2022AJ....163..218H} for Lalande 21185, and \cite{2019arXiv190604644T,2023AJ....166..260B} for Wolf 359.}
\end{figure}

Planetary habitability is one possible axis on which to prioritize target selection for SETI observations \citep{2001ARA&A..39..511T,2004NewAR..48.1543T,2003ApJS..145..181T}. Stars with planets located within HZ are considered prime candidates which are possible to maintain liquid water on planetary surfaces, increasing the likelihood of life and, consequently, potential technosignatures. We also estimate the planetary habitabilities around these stars using the stellar flux HZ model in \cite{2013ApJ...765..131K} and the ``Habitable Zone for Complex Life (HZCL)'' model in \cite{2019ApJ...878...19S}, in which the stellar effective temperatures, stellar luminosity, \ce{H2O} and \ce{CO2} absorption coefficients, as well as the planetary atmospheric \ce{CO2} pressure are taken in to consideration. Figure \ref{fig:habitablezone} illustrates the boundaries of the HZ.  
We can see that Upsilon Andromedae A d and Gliese 581 d are near the edge of maximum greenhouse, and Gliese 581 g, if it exists, falls in the HZCL. 

Most of flare stars are M dwarfs \citep{1989A&A...217..187P,2000eaa..bookE1866G}, and flare event records have also been reported for some selected targets, including Barnard's star \citep{2006PASP..118..227P,2020AJ....160..237F}, Ross 128 \citep{1972IBVS..707....1L}, Lalande 21185 \citep{1995ApJ...450..392S,2015A&A...581A..28P} and Wolf 359 \citep{1983Ap&SS..95..235G,1995ApJ...451..795R,1995ApJ...450..392S,2007MmSAI..78..258L}. Stellar magnetic activities, especially stellar flare activities, can significantly impact the radiation environment encountered by planets, making it an essential role in promoting or destroying planetary habitabilities. Generally, intense stellar flares and radiation bursts are viewed negatively, as they may severely erode planetary atmospheres or expose surfaces to sterilizing radiation \citep{2017NatSR...714141A,2017ApJ...844L..13G}, consequently threaten the habitability for complex life-forms. It is proposed that, on the other hands, moderate, or even intense, energetic eruptive stellar activity might drive the planetary atmosphere dynamics via interactive photochemistry, thereby affect the climate and atmospheric evolution of the planet \citep{2022A&A...667A..15K,2024MNRAS.532.4436B,2025AJ....170...40C}. Moerover, the synthesis of some vital biochemical products for complex life, such as Vitamin D \citep{2023MNRAS.522.1411S}, amino acids \citep{2013AdSpR..51.2235S}, and ribonucleic acid \citep{2009Natur.459..239P,2016AsBio..16...68R,2018SciA....4.3302R}, can be triggered under the condition of ultraviolet (UV) light from stellar radiation and flare. 

\subsubsection{SETI/METI Search for the Targets}
Some of the observation targets are also listed in some SETI/METI related project. A digital radio signal named ``A Message from Earth'' has been sent towards Gliese 581, and a message call ``Cosmic call 2'' has been sent towords 55 Cancri\footnote{\url{https://www.plover.com/misc/Dumas-Dutil/messages.pdf}}, respectively, by the Yevpatoria RT-70 radio telescope. Gliese 581 is targeted by the first Very Long Baseline Interferometric (VLBI) SETI experiment, while no candidate was found \citep{2012AJ....144...38R}. A anomalous unpolarized radio emission was detected during the observation of Ross 128 \footnote{\url{https://phl.upr.edu/library/notes/ross128}}, which is confirmed to be RFI by the follow-up observation \citep{2017arXiv171008404E,2017ApJ...849..104E}. Previous FAST SETI observations also attempt to search radio technosignature from Barnard's star \citep{2023AJ....166..190T} and Wolf 359 \citep{2025AJ....169..217L}.

\subsection{Strategy}\label{subsec:Strategy}
Currently, there are 10 categories of observation mode\footnote{\url{https://fast.bao.ac.cn/cms/article/24/}} and 3 digital backend\footnote{\url{https://fast.bao.ac.cn/cms/article/26/}} for FAST, working at the frequency range of 1.05 GHz-1.45 GHz with full-polarization measurements from two linear feeds. Previously, drifting scan \citep{2020ApJ...891..174Z}, tracking \citep{2022AJ....164..160T,2023AJ....166..190T,2023AJ....165..132L,2025AJ....169..217L}, and multibeamOTF \citep{2023AJ....166..245H} modes are employed in the FAST SETI observation. To make full use of the layout of the 19-beam receiver and the digital backends of FAST, as well as to improve the RFI rejection process by instrumental calibration, we employ position switching observation mode and record data with both psr backend and SETI backend. For flux and polarization calibration, we also inject linearly polarized noise diode signals with temperature of $\sim12.5\,\mathrm{K}$ for $\sim5\,\mathrm{s}$ at the beginning of the first and last ON and OFF observations, respectively.
\subsubsection{Multibeam Position Switching Mode}\label{subsubsec:MBOM}
The on-off strategy, or generally, position switching, is a standard observation mode for targeted SETI and spectral line studies, which implements repetitive paired cycles of telescope movements between on-source and adjacent off-source positions at matched airmass. For SETI applications, artificial technosignatures are expected to manifest exclusively in on-source observations while remaining statistically absent in off-source reference data, providing critical discrimination against terrestrial RFI and instrumental artifacts \citep{2017ApJ...849..104E}. For large single-dish radio telescopes, multibeam observations are often used in large sky serveys for high efficiency \citep{1996PASA...13..243S,2006ApJ...637..446C,2018IMMag..19..112L}. 

Multibeam strategy has also been implemented in targeted SETI observations using FAST's 19-beam receiver, significantly enhancing temporal efficiency and RFI mitigation capabilities \citep{2022AJ....164..160T,2023AJ....165..132L}. Spatial beam diversity provides multiple concurrent reference positions, enabling advanced RFI discrimination through coverage pattern analysis. Nevertheless, beam variations in gain, aperture efficiency, system temperature, and polarization leakage can compromise RFI rejection validity without rigorous per-beam calibration. 

The position switching observation is also a fundamental calibration technique essential for radio astronomy. By pointing the telescope at the target (on) and then at a nearby empty patch of sky (off), the background noise the atmosphere and the instrument itself can be measured and subtracted, which enable us to calculate the system's true sensitivity. The observation modes we use in this work are OnOff mode and PhaseReferencing mode, the observation details of the sources are listed in Table \ref{table:Target_obs}. The basic position switching for FAST is OnOff mode, where the on-source point and off-source point are observed with equal time, while PhaseReferencing, one of the OnOff mode extension, enables us to set the observation time and the positions for the on-source point and off-source point separately. 
\begin{deluxetable}{lccccc}
\tablecaption{Observation Details of the Targets in This Work.\label{table:Target_obs}}
\tablehead{
\colhead{Target} & 
\colhead{Observation Date} & 
\colhead{Observation Mode} & 
\colhead{OFF  R.A. (J2000.0)} & 
\colhead{OFF  Decl. (J2000.0)} & 
\colhead{Off-source Beam} 
}
\startdata
3C286            & 2024-11-15 & OnOff & 13:31:08.28  & +30:00:32.9  & --   \\
Barnard's Star   & 2024-11-15 & OnOff & 17:57:47.67  & +04:14:16.7  & --   \\
Ross 128         & 2024-11-15 & OnOff & 11:47:45.02  & +00:17:57.4  & --   \\
3C48             & 2025-07-09 & PhaseReferencing & 01:38:08.58  & +33:19:32.7& 12   \\
3C147       & 2025-07-09 & PhaseReferencing & 05:43:11.56  & +50:01:04.8& 12   \\
3C286            & 2025-07-09 & PhaseReferencing & 13:32:01.68  & +30:30:34.8& 14   \\
Gliese 581       & 2025-07-09 & PhaseReferencing & 15:20:13.25  & $-$07:43:18.4& 14   \\
Upsilon Andromedae A   & 2025-07-09 & PhaseReferencing & 01:37:49.17  & +41:24:21.5& 14   \\
55 Cancri A            & 2025-07-09 & PhaseReferencing & 08:53:28.07  & +28:19:52.8& 14   \\
Lalande 21185          & 2025-07-09 & PhaseReferencing & 11:04:17.03  & +35:58:13.4& 14   \\
Wolf 359               & 2025-07-09 & PhaseReferencing & 10:56:51.93  & +07:10:50.6& 12   \\
\enddata
\end{deluxetable}

\begin{figure}[htbp]
    \centering
    \includegraphics[width=0.8\textwidth]{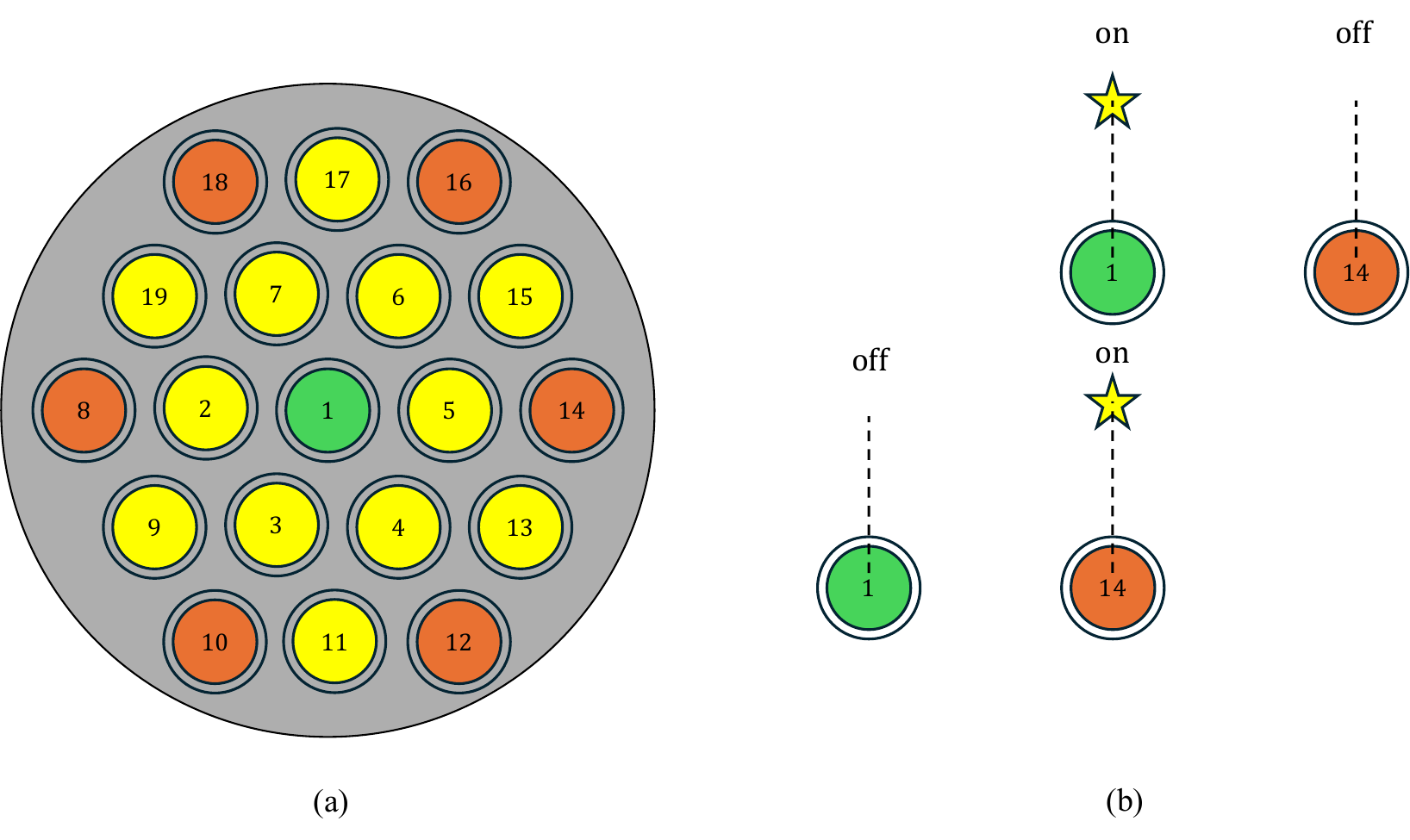}
    \caption{A simple diagram for (a) the multibeam layout of FAST and (b) an example of multibeam position switching method. In the on-source observations, the central beam 1 points at the target, serving as ``On''. In the off-source, one of the edge beam 14 switchs to the location pointing at the target, serving as ``On'', while beam 1 serves as ``Off''.}
    \label{fig:MBPoSw}
\end{figure}

Our observation strategy employs a on-off cycle with 3 repetitions for each target. In the On-Off observations, we set the observation time for each On/Off as 6 minutes. The off-source position is defined by a $0.5^{\circ}$ offset in declination from the on-source target. The angular separation is significantly larger than the Full-Width Half-Maximum (FWHM) beamwidth of FAST at 1.4 GHz (approximately $2.9'$), ensuring that the target source is well outside the primary beam of FAST during off-source measurements. Such configuration guarantees that any signals detected in the off-source across multiple beams can be identified as RFI. While the PhaseReferencing observations consist of 6-minute on-source time and 30-second off-source time. By carefully setting the off-source coordinates, the telescope's pointing is optimized such that when the central beam is positioned on the off-source location, a designated edge beam simultaneously points at the on-source target (See Figure \ref{fig:MBPoSw} (b)). 
This interleaved approach allows us to conduct a necessary off-source observation for one beam while simultaneously performing an on-source observation for another, effectively increasing the on-target integration time within a given observing window.

\subsubsection{FAST Digital Backend}\label{subsubsec:Backend}
There are three kinds of backends for FAST : psr, spectral (spec) and SETI, which can be in different combinations with single or multiple backend configuration. Since the SETI and spec backends use the same computing resources, the SETI and spec data cannot be recorded simultaneously. For sake of high temporal resolution and high-frequency resolution, we select SETI backend with $\sim10\,\mathrm{s}$ sampling time and 7.5 Hz frequency resoltion, and psr backend with $49.152 \,\mathrm{\mu s}$ sampling time and 0.122 MHz frequency resoltion. The details of FAST multibeam digital backend are described in \cite{2020ApJ...891..174Z,2020RAA....20...78L}. After the observations, we can obtain psrfits files with psr backend and sdfits files with SETI backend for each target, and the FITS files are archived for each beam, each FITS file corresponds to a different sample time segment of the observation. All these FITS files are merged along time axis in order for each beam, and converted into Filterbank files containing two dimensional time-frequency power spectra of the four polarization channels: XX, YY, XY and YX, which are derived from self-correlation and cross correlation of the data in two orthogonal linear polarization directions X and Y. The psrfits files are converted into Filterbank files by digifil in DSPSR \citep{2011PASA...28....1V}, and the sdfits files are converted into Filterbank files by filterbank package in PRESTO \citep{2011ascl.soft07017R}. All these Filterbank files are accessible to the Blimpy package \citep{2019JOSS....4.1554P}.

\section{Data Process and Analysis}\label{sec:DataAnalysis}
Given that the duration of noise diode signals is shorter than the backend sampling time in SETI observations, we explicitly subtract these signals from PSRFITS data based on their precise injection timestamps. The noise diode injections correct amplitude mismatches between the two linear polarization feeds. Digitized outputs are converted to antenna temperature $T_a$ using the noise diode calibration report, with flux density scaling achieved through observations of calibrators. We flag RFI channels by firstly identifying and masking significant outliers in the time-averaged spectrum, defined as channels with intensities deviating from the global median and median absolute deviation (MAD). Subsequently, a smooth spectral baseline is estimated by applying a median filter to the remaining unflagged channels. This baseline is used to identify weaker RFI by flagging any channels where the residual spectrum exceeds the fluctuation threshold (See Appendix \ref{subsec:appendix_RFIFlagging} for details).

Narrowband drifting signal search can be done by TurboSETI, a python-cython package to search for narrowband signals via Taylor-tree de-doppler algorithm \citep{1974A&AS...15..367T,2013ApJ...767...94S,2017ApJ...849..104E,2019ascl.soft06006E}. Narrowband continuous signals of extraterrestrial origin should exhibit Doppler frequency drift at a rate $\mathrm{d}\nu/\mathrm{d}t$ due to the relative motion in the line of sight \citep{2022ApJ...938....1L}. Over short observational timescales, this drift rate is effectively constant, producing a linear frequency variation. TurboSETI amis to search such drifting signal above the signal-to-noise ratio (SNR) threshold within a given maximum drift rate (MDR). Following many previous targeted SETI observations, as well as the MDR for the each planet calculated based on \cite{2022ApJ...938....1L}, we set the SNR threshold as 10 for all targets, while the MDR used for each target is listed in Appendix \ref{subsec:appendix_MDR}. 

We employ the blipss pipeline \citep{2023AJ....165..255S} to search for periodic pulsed signals from the psr backend data, utilizing a Fast Folding Algorithm (FFA) for channel-wide periodicity detection. The parameters used in the search are listed in Table \ref{tab:blipss_parameters}.
\begin{table}[htbp]
\centering
\caption{Parameter values used in blipss search in this work.}
\label{tab:blipss_parameters}
\begin{tabular}{@{}lc@{}}
\toprule
Parameter & Value \\ 
\midrule
Running median width, $W_{\mathrm{med}}$ & 12 s \\
Range of trial periods $P$ & 0.12-100 s \\
Pulse duty cycle resolution & 10\% \\
Range of trial duty cycles $\delta$ & 10\%-50\% \\
S/N threshold for ON pointings & 8 \\
S/N threshold for OFF pointings & 6 \\
\bottomrule
\end{tabular}
\end{table}
While most parameters adopt the default configurations from \cite{2023AJ....165..255S}, the minimum trial period is explicitly set to 0.12 s, which still satisfies the fundamental constrain $P_{\min}\geqslant N_\mathrm{bins}t_\mathrm{samp}$ with $N_\mathrm{bins}=10$ being the minimum bin number across a folding period. Valid candidates must exhibit statistically significant folded pulse profiles detected exclusively in on-source observations.

To avoid ambiguity and maintain consistency, we define signals detected above signal-to-noise ratio threshold as $\emph{hit}$. A set of $\emph{hits}$ present in ``On'' observations with relatively noticeable continuities in time or frequency are grouped into an $\emph{event}$. Event that contain no hits in all ``Off'' observations are defined as $\emph{candidate}$. For narrowband drifting signal, the definations of event $\epsilon_\mathrm{NB}$ and candidate $\mathcal{C}_\mathrm{NB} $ can be formulated by Equation (\ref{eq:event_NB}) and (\ref{eq:candidate_NB}) via hits $h$ \citep{2021AJ....161..286T}:
\begin{equation}
  \epsilon_\mathrm{NB} = \{h\in \mathrm{On_k}:\nu_0-\dot{\nu}_0\tau_\mathrm{obs}\leqslant \nu_h\leqslant\nu_0+\dot{\nu}_0\tau_\mathrm{obs}\},\label{eq:event_NB}
\end{equation}
\begin{equation}
  \mathcal{C}_\mathrm{NB}  = \{h\in \mathrm{On_k}:h\notin\mathrm{Off}\cap\nu_0-\dot{\nu}_0\tau_\mathrm{obs}\leqslant \nu_h\leqslant\nu_0+\dot{\nu}_0\tau_\mathrm{obs}\},\label{eq:candidate_NB}
\end{equation}
where $\nu_h$ is the frequency of a hit in the kth ``On'' observation, $\nu_0$ and $\dot{\nu}_0$ are the central frequency and corresponding drift, respectively. For periodic signal, event $\epsilon_\mathrm{P}$ and candidate $\mathcal{C}_\mathrm{P} $ can be defined by
\begin{equation}
  \epsilon_\mathrm{P} = \{h\in \mathrm{On_k}: P_{\min}\leqslant P_h\leqslant P_{\max}\},\label{eq:event_P}
\end{equation}
\begin{equation}
  \mathcal{C}_\mathrm{P}  = \{h\in \mathrm{On_k}:h \notin \mathrm{Off}\cap P_{\min}\leqslant P_h\leqslant P_{\max} \},\label{eq:candidate_P}
\end{equation}
where $P_h$ refers to the period of the hits' appearance in the frequency channel.

\section{Results}\label{sec:Results}
\subsection{Narrowband Signal Search}\label{subsec:NBSearch}
\begin{figure}[htpb]
  \centering
  \includegraphics[width=\textwidth]{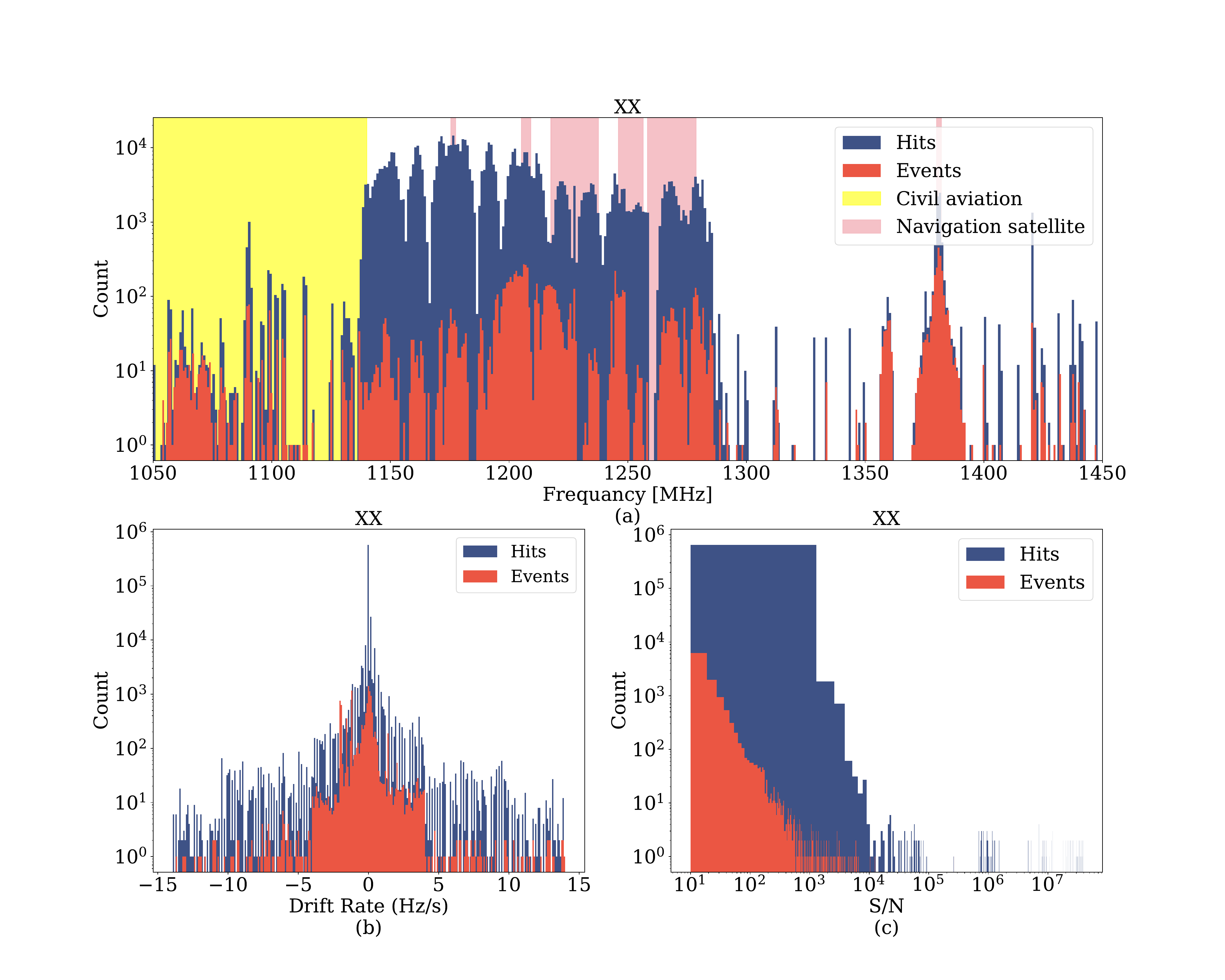}
  \caption{\label{fig:turboseti_statistics_XX} Histograms of the distributions of frequency, drift rate and signal-to-noise ratio in XX channel. The frequency bands of some known interference sources are displayed on the frequency panel. }
\end{figure}

\begin{figure}[htpb]
  \centering
  \includegraphics[width=\textwidth]{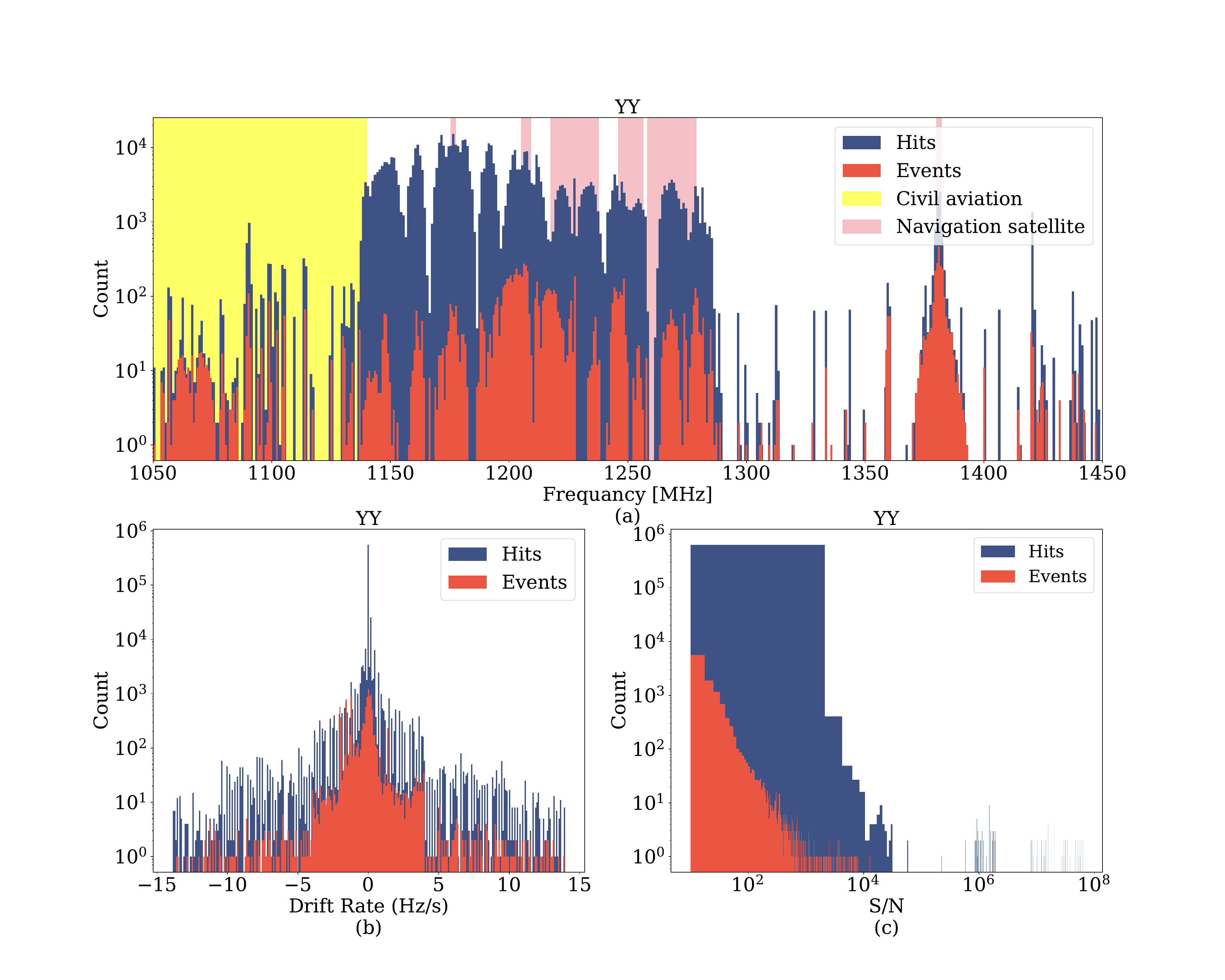}
  \caption{\label{fig:turboseti_statistics_YY} Similar histograms of the distributions of frequency, drift rate and signal-to-noise ratio in YY channel.}
\end{figure}

\begin{figure}[htpb]
  \centering
  \includegraphics[width=\textwidth]{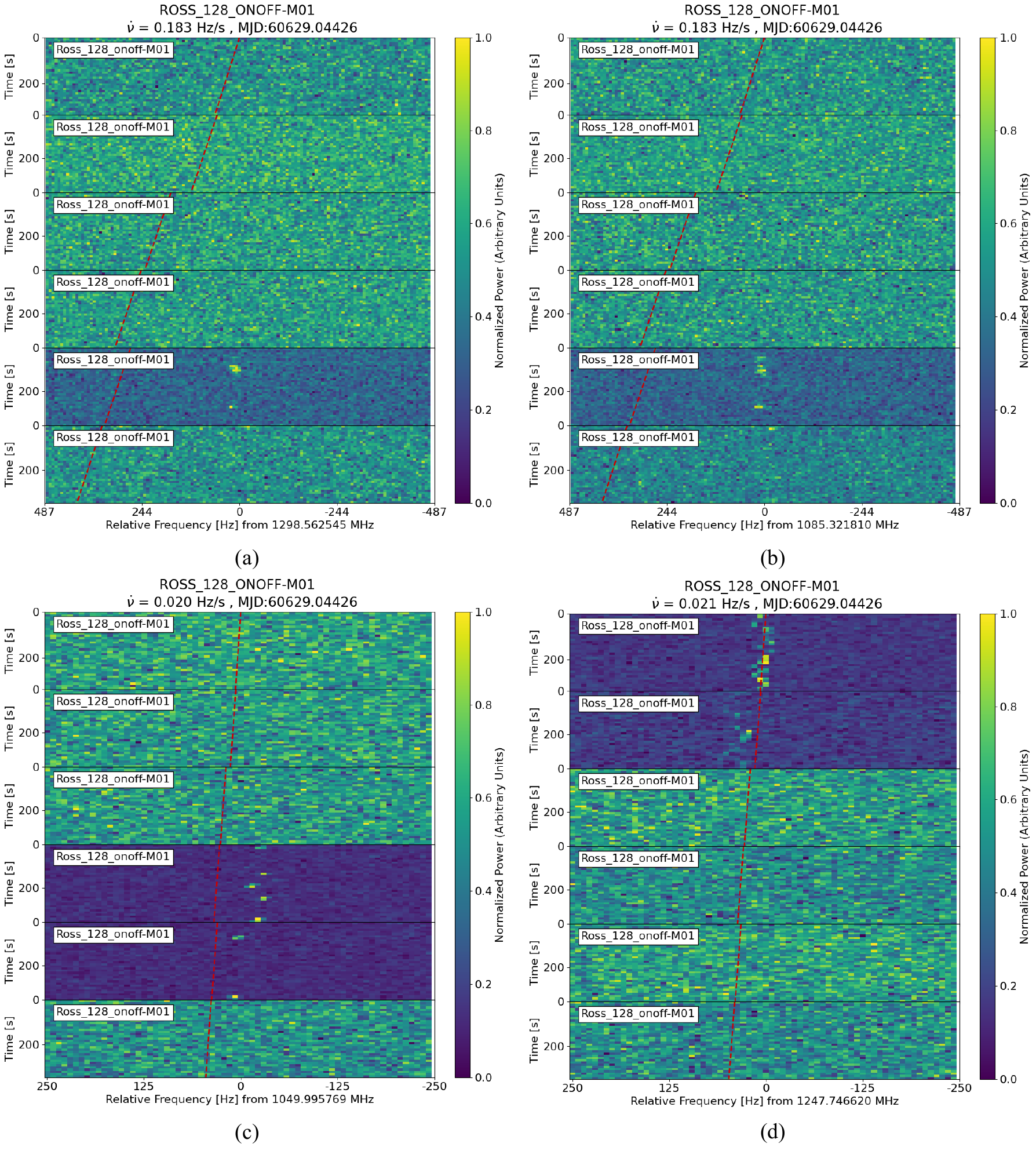}
  \caption{\label{fig:events_example} Examples of RFIs that only occupy several time-frequency pixels. Panel (a) is an event that appears in 1298.562058 MHz, which is not the categorized as RFI frequency ranges. Panel (b) is an event that appears in 1085.321323 MHz, which fall in the frequency of civil aviations. Panel (c) and (d) are evnets that have similar patterns with the events in Panel (a) and (b), but also appear in off-source observations.}
\end{figure}

In the narrowband drifting signal search, we found 653804 hits in XX polarization channel and 634505 hits in YY polarization channel after running turboSETI. Among these hits, 11537 events in XX and 11491 events in YY are detected in the on-source observations. The distributions of frequency, drift rate, and signal-to-noise ratio are illustrated in Figure \ref{fig:turboseti_statistics_XX} and \ref{fig:turboseti_statistics_YY}, and some known RFI sources within the observation frequency ranges \citep{2021RAA....21...18W} are also shown in the plots. 
A considerable amount of hits fall in the frequency ranges of civil aviations (1.40\% for XX and 1.88\% for YY, respectively) and navigation satellites (24.42\% for XX and 25.45\% for YY, respectively), implying that about 30\% RFIs may come from these knows sources. 255 events in XX and 252 events in YY fall near the frequencies of the clock oscillators used by the Roach 2 FPGA board, wihch can be calculated by the linear combinations ofthe nominal frequencies 33.3333 and 125.00 MHz. We also reexamine the selected events by visual inspection of the time-frequency spectra. Almost all of the selected events are false positives mentioned in \cite{2022AJ....164..160T}, which can be directly excluded. We also find some events that only occupy several time-frequency pixels, instead of spanning the full on-source time. Figure \ref{fig:events_example}(a) shows a representative event, and we also find 29 events in XX and 12 events in YY with similar morphological feature that only appear during the third on-source observation. Among these events, 27 in XX and 11 in YY of them are within frequency ranges of known RFI sources, and Figure \ref{fig:events_example} (b) shows one such case. These events can be excluded by examining other events with similar morphological feature that also appear during the off-source observations, as illustrated in Figure \ref{fig:events_example} (c) and (d), and we therefore incline to classify them as RFIs rather than potential technosignatures.

\subsection{Periodic Pulsed Signal Search}\label{subsec:PPSearch}
For the OnOff observations in 2024 November 15, we divide observation data into 3 On and Off 6-minute series respectively and carry out FFA search in single beam. For the PhaseReferencing observations in 2025 July 9, the central and reference beams alternate roles as asymmetric On and Off observation, we can conduct FFA search for the 6-minute and 30-second series composed by the two beams. Following the defination in \cite{2023AJ....165..255S}, we classify $F_n$ as any candidate detected in exactly $n$ on-source scans and absent from all off-source scans. The results of the FFA search are listed in Table \ref{tab:blipss_candidates}, including 1830177 $F_1$ candidates and 585178 $F_2$ candidates in total. We cautiously extrapolate that the candidate differences in number may be caused by the different RFI environment at FAST, the diverse gain responses in each beam, and  inequality in On/Off time setting of the observation modes (See Appendix \ref{sec:appendix_cand_stat} for detailed statistical attribution).

\begin{deluxetable}{lccccc}
\tablecaption{Candidate Statistics for Target Sources before and after filtering.}\label{tab:blipss_candidates}
\tablehead{
\colhead{Source} & \colhead{$F_1$} & \colhead{$F_2$} &
\colhead{$F_{1,\mathrm{filtered}}$} & \colhead{$F_{2,\mathrm{filtered}}$} &
\colhead{$N_{\mathrm{event}}$}
}
\startdata
Barnard's Star         & 6283   & 0      & 3    & 0    & 55135   \\
Ross 128               & 7603   & 0      & 0    & 0    & 59777   \\
55 Cancri A            & 898530 & 276513 & 0    & 0    & 5155016 \\
Gliese 581             & 351232 & 130461 & 0    & 0    & 5211503 \\
Lalande 21185          & 218800 & 81791  & 760  & 363  & 987255  \\
Upsilon Andromedae A   & 206928 & 69961  & 8    & 7    & 1329432 \\
Wolf 359               & 192133 & 72336  & 9405 & 2919 & 1186105 \\
\hline
Total                  & 1881509& 631062 & 10176& 3289 & 13984223\\
\enddata
\end{deluxetable}

\begin{figure}[htpb]
  \centering
  \includegraphics[width=\textwidth]{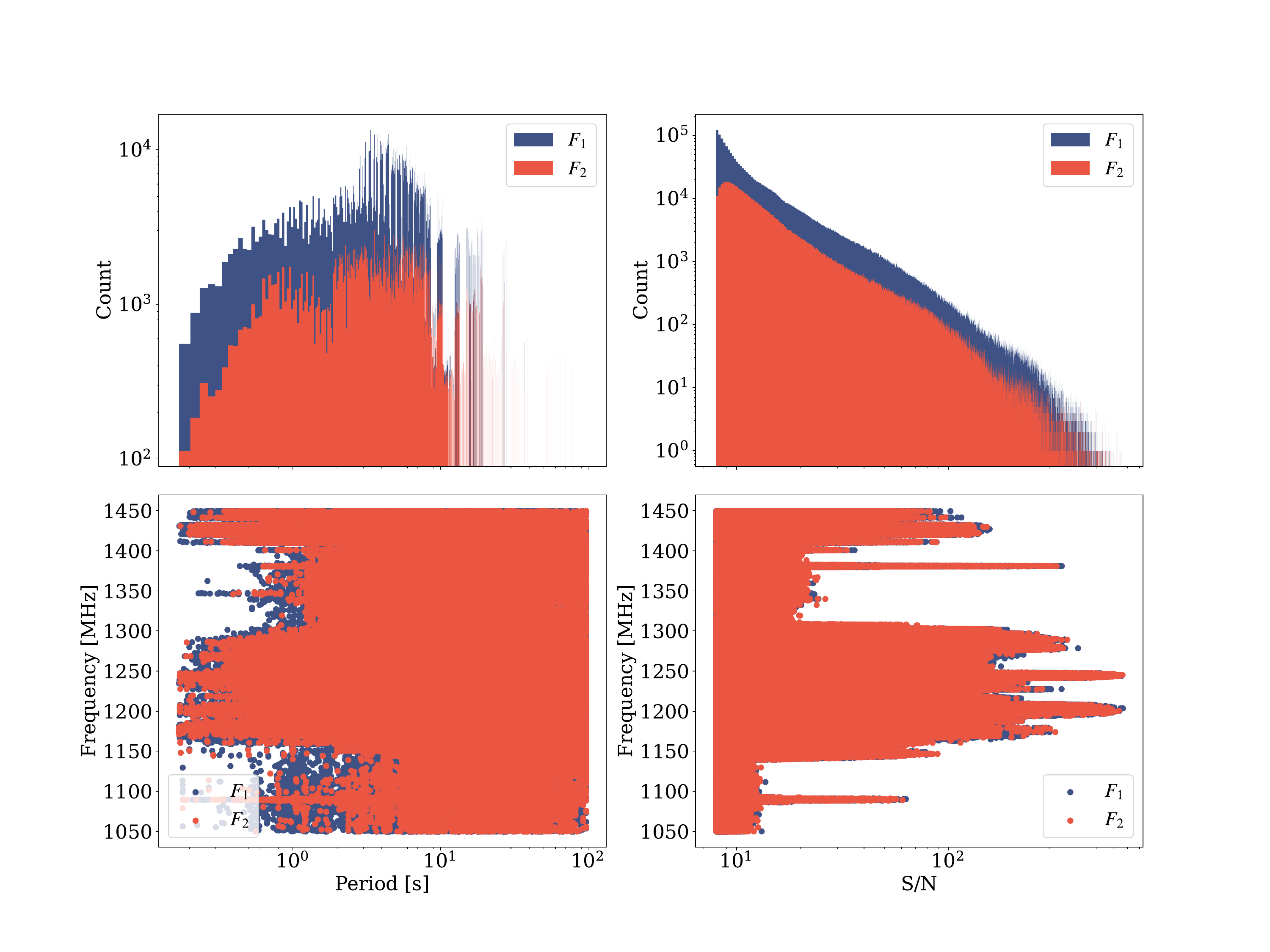}
  \caption{\label{fig:blipss_statistics} Statistical distribution of $F_1$ (blue) and $F_2$ (red) in the search.Top left panel: histogram of the candidate periods. Top right panel: histogram of the signal-to-noise ratio of the candidates. Bottom left panel: scatter of candidates in $\nu-P$ plane. Bottom right panel: scatter of candidates in $\nu-\mathrm{S/N}$ plane.}
\end{figure}

\begin{figure}[htpb]
  \centering
  \includegraphics[width=\textwidth]{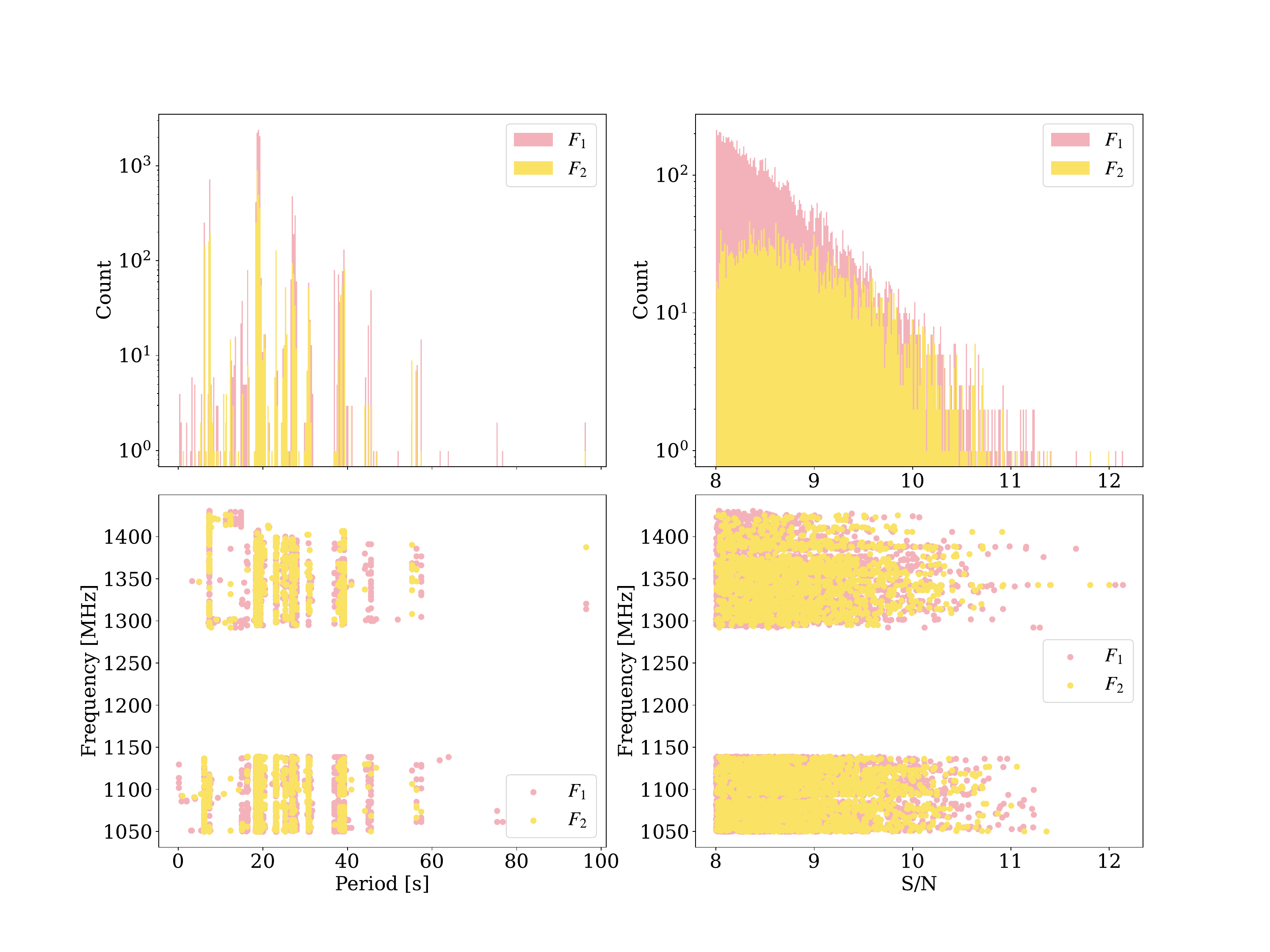}
  \caption{\label{fig:blipss_statistics_select} Statistical distribution of $F_1$ (pink) and $F_2$ (yellow) in the search silimar to Figure \ref{fig:blipss_statistics} but after filtering.}
\end{figure}

\begin{figure}[htpb]
  \centering
  \includegraphics[width=0.8\textwidth]{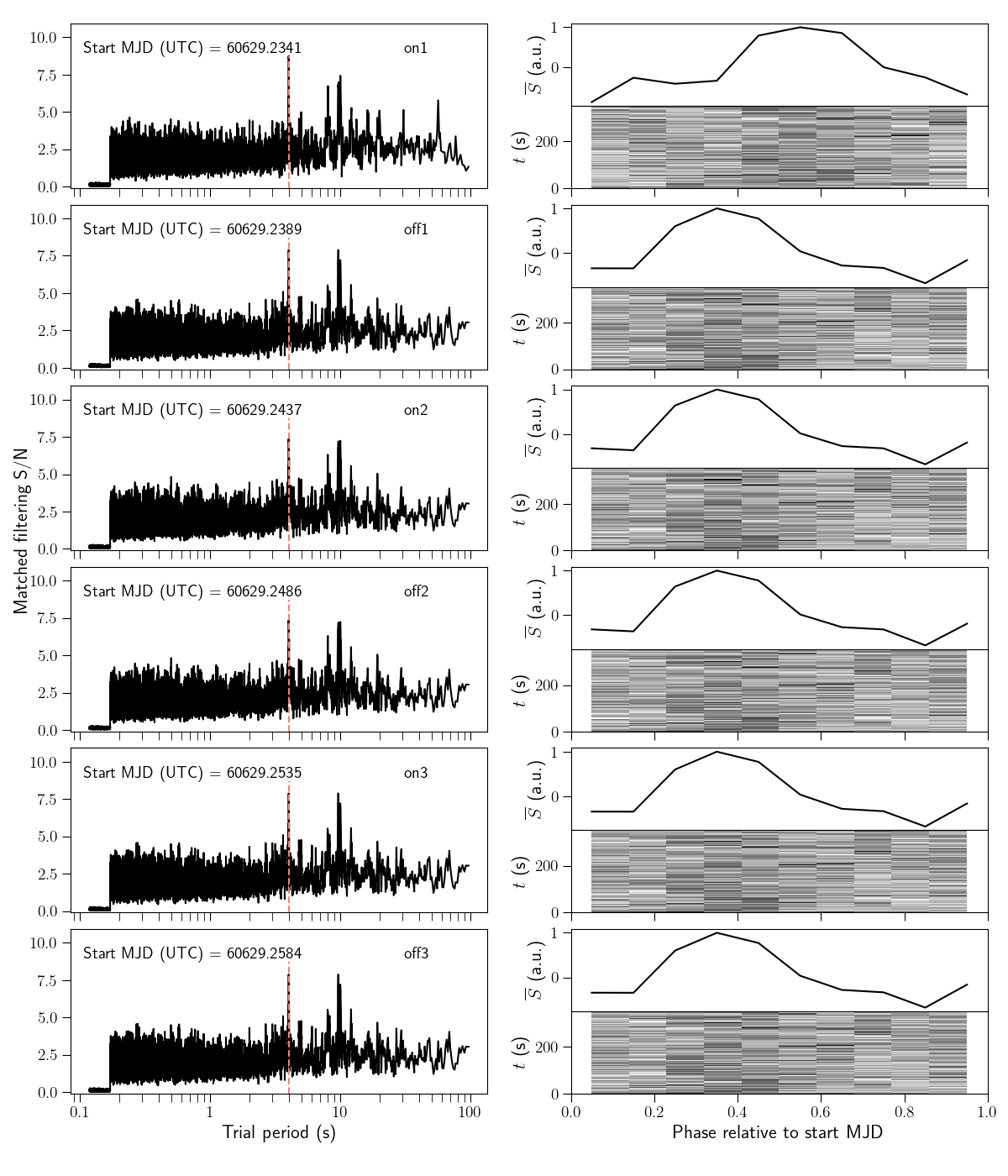}
  \caption{\label{fig:Barnard_star_blipss_candidate} Plot for the candidate in the in 1089.111328 MHz channel during the observation of Barnard's star. Left panels are the periodograms generated from the time-series. Right panels show the illustrate phase-time diagrams (bottom subplots) and folded profiles (top subplots) of candidates at $P=4.00106$ s. The peak of the periodogram and the folded profile are detected in the first on-source scan. }
\end{figure}

Figure \ref{fig:blipss_statistics} exhibits the statistical distribution of candidates in $\nu-P$ and $\nu-\mathrm{S/N}$ diagrams. The statistics reveals that most of the candidates are predominantly populated by RFI, wihch is evidenced by the nearly identical statistical distributions of $F_1$ and $F_2$ candidates across both period and signal-to-noise ratio, strongly suggesting a common origin. Crucially, both populations densely populate almost the whole frequency-period parameter space, and cluster within narrow, persistent radio frequency channels. 
To substantially reduce the candidates, we entirely discard candidates from frequency channels with severe persistent RFI within 1140 MHz to 1290 MHz. Specifically, for the remaining candidates, we perform a comparison with candidates detected during ``Off'' observations. A candidate should be excluded as RFI if its relative tolerance agrees within 1\% in period and 5\% in signal-to-noise ratio (S/N) relative to events found in the same frequency channel during an ``Off'' scan, assuming they share a common terrestrial origin. Figure \ref{fig:blipss_statistics_select} shows the statistical distribution of the candidate population after this comprehensive filtering process, the candidate numbers for each target are also listed in \ref{tab:blipss_candidates}. 
For these remaining candidates, we generate the periodogram, phase-time diagrams and folded profiles to visualize them. Figure \ref{fig:Barnard_star_blipss_candidate} plots an example of $F_1$ candidate with $P=4.00106$ s in 1089.111328 MHz channel, and the other two remaining candidates are also appear around the period 4 s in this channel. Since the shapes of the periodograms and folded profiles of these six observations are very similar, with only the phases during the on-source observation period being different in the hump of the profiles, we tend to believe that the signals in the on-source scan and other are likely to come from the same source. To quantify these feature, we carry out Kruskal-Wallis test for the trial periods and S/N of periodograms as well as the folded profiles. For $g$ independent samples with size $(n_1,n_2,\dots,n_g)$, the Kruskal-Wallis statistic can be expressed as \citep{1952JST...47...583}

\begin{equation}
  H={\frac {12}{N(N+1)}}\sum _{i=1}^{g}\frac{R_{i}^{2}}{n_{i}}- 3(N+1),
\end{equation}
where $N=\sum n_i$ is the total sample size and $R_{i}$ is the sum of ranks in the $i$-th group. The null hypothesis $H_0$ states that all groups are drawn from the same population. The p-value of the test is obtained from the survival function of the chi-square distribution evaluated at $H$. If p-value exceeds the significance level $\alpha$, we fail to reject the null hypothesis $H_0$, which means there is insufficient evidence to conclude that the groups differ. By Kruskal-Wallis test and visual inspections for these remaining candidates, we reject all of them as false positives.

\begin{figure}[htpb]
  \centering
  \includegraphics[width=\textwidth]{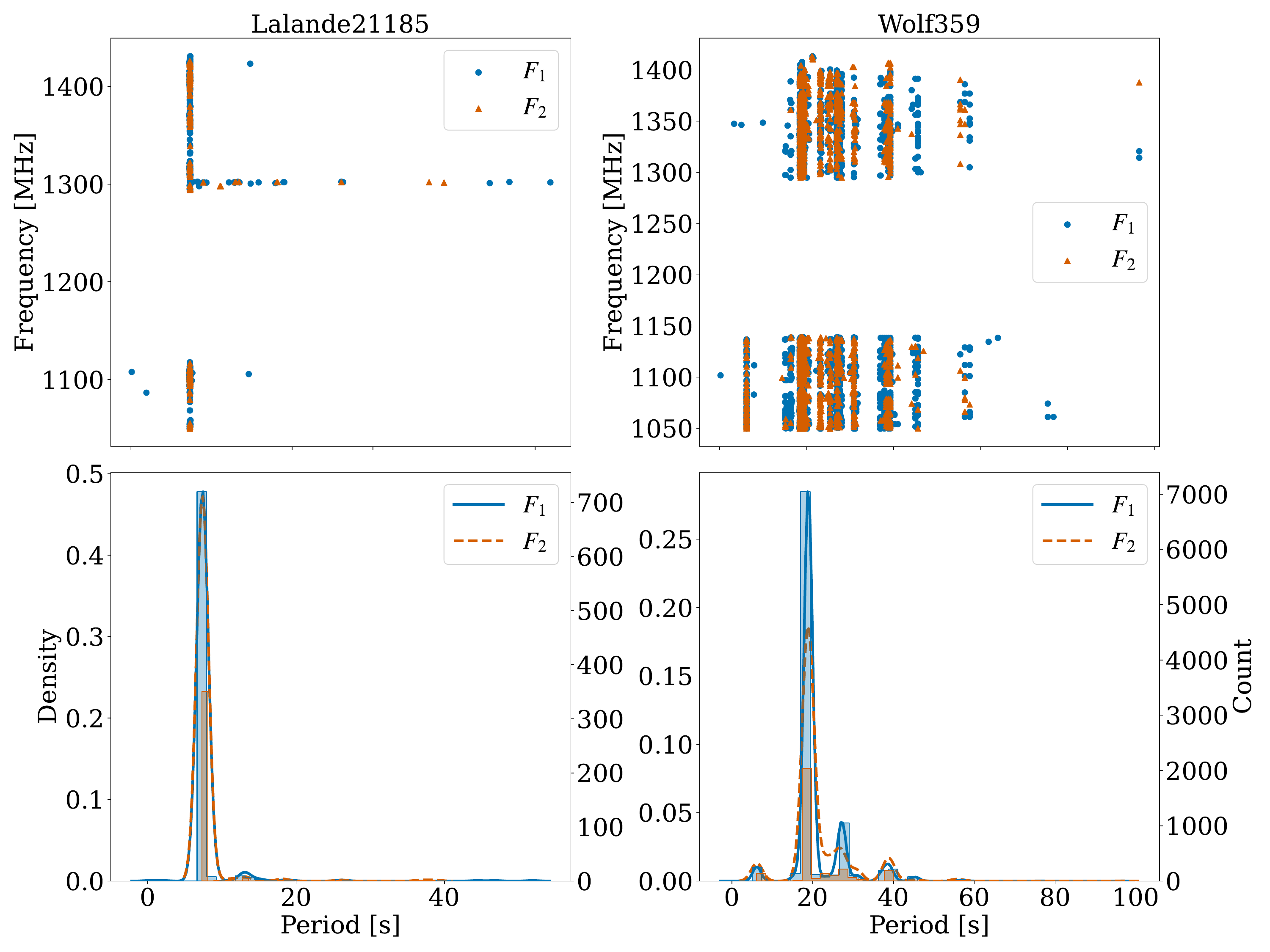}
  \caption{\label{fig:cand_prob_stat} Top row panels: Scatter plots of the remaining candidates in $\nu-P$ plane for Lalande 21185 (left) and Wolf 359 (right). Bottom row panels: histogram and corresponding probability density profiles of the candidates.}
\end{figure}

A subset of remaining candidates still clustered at specific periods after applying the filtering, wihch means broadband pulses may appear in the observation. Since most of the remaining candidates are from Wolf 359 and Lalande 21185, and these candidates appear at in certain periods across all frequency channels (See the top panels of Figure \ref{fig:cand_prob_stat}), we generated phase-resolved spectra for these two sources by folding at the most probable period identified from the candidates' probability density and the distribution. For the candidates from Lalande 21185, as illustrated in the bottom left panel of Figure \ref{fig:cand_prob_stat}, the peak of the probability density is at $P=7.45293 \ \mathrm{s}$ and almost all candidates are concentrated around this period.  The candidate distribution for Wolf 359 is more complex, except the primary peak at $P_\mathrm{main}=18.76658 \ \mathrm{s}$, there are also three secondary peaks at $P_\mathrm{sub1}=6.16899 \ \mathrm{s}$, $P_\mathrm{sub2}=27.18886 \ \mathrm{s}$, and $P_\mathrm{sub3}=38.53254 \ \mathrm{s}$, respectively.  We note a near-rational relationship between these periods, approximated by $[P_\mathrm{sub1}/P_\mathrm{main}, P_\mathrm{sub1}/P_\mathrm{sub2}, P_\mathrm{sub1}/P_\mathrm{sub3}]\sim[1/3, 2/9, 1/6]$ for the four peaks in the bottom right panel. Visual inspection of these phase-resolved spectra yields no evidence of periodic signal or dispersed pulse in the corresponding frequency channel (See Appendix \ref{sec:appendix_PRS_analysis} for the example phase-resolved spectra). Consequently, we attribute these remaining candidates to periodic interference in the noise. The observation of Lalande 21185 may be affected by a source of simple periodic interference at $P=7.45293 \ \mathrm{s}$ in such noise fluctuations, whereas for the observation of Wolf 359, the presence of near subharmonic and fractional harmonic relation for the four peaks suggests a complex periodic noise interference source with a fundamental period near $P_\mathrm{sub1}$, and the candidate difference in number suggests complex amplitude modulations of the interference.

\section{Discussion}\label{sec:Discussion}
\subsection{Sensitivity}\label{subsec:Sensitivity}
The sensitivity of a radio observation can be determined by system equivalent flux density \citep{2013tra..book.....W,2017isra.book.....T}
\begin{equation}
  \mathrm{SEFD}=\frac{2k_\mathrm{B}T_\mathrm{sys}}{A_\mathrm{eff}},
  \label{SEFD}
\end{equation}
where $k_\mathrm{B}$ is the Boltzmann constant, $T_\mathrm{sys}$ is the system temperature, and $A_\mathrm{eff}$ is the effective collecting area. The sensitivity $A_\mathrm{eff}/T_\mathrm{sys}$ of FAST L-band 19 beam receiver is $\sim 2000 \, \mathrm{m^{2} \, K^{-1}}$ \citep{2011IJMPD..20..989N,2016RaSc...51.1060L,2019SCPMA..6259502J}. For narrowband signal detection (i.e., the signal bandwidth is narrower or equal to the observing spectral resolution), the minimum detectable flux density $S_{\min}$ can be given by \citep{2017ApJ...849..104E}
\begin{equation}
  S_{\min,\mathrm{N}}=\mathrm{SNR}_{\min}\mathrm{SEFD}\sqrt{\frac{\delta \nu_{\mathrm{ch}}}{n_\mathrm{pol}\tau_\mathrm{obs}}},
  \label{NBS_min}
\end{equation}
where $\mathrm{SNR}_{\min}$ is the signal-to-noise ratio threshold, $\tau_\mathrm{obs}$ is the effective observing duration, $\delta \nu_{\mathrm{ch}}$ is the frequency channel bandwidth and $n_\mathrm{pol}$ represents the number of polarization channels of the telescope. While for channel-width periodic signal detection, $S_{\min}$ should be \citep{2020MNRAS.497.4654M,2023AJ....165..255S}
\begin{equation}
  S_{\min,\mathrm{P}}=\mathrm{SNR}_{\min}\frac{\mathrm{SEFD}}{\mathcal{E}}\sqrt{\frac{\delta \nu_{\mathrm{ch}}}{n_\mathrm{pol}\tau_\mathrm{obs}}}\sqrt{\frac{\delta}{1-\delta}},
  \label{BBS_min}
\end{equation}
where $\mathcal{E}=0.93$ is a function of effective pulse duty cycle $\delta=t_{\mathrm{pulse}}/P$ for practical use. For $\mathrm{SNR}_{\min}=10$ and $\tau_{\mathrm{obs}}\sim 6 \,\mathrm{min}$, the minimum detectable flux density for narrowband signal $S_{\min,\mathrm{N}}$ is $0.99 \,\mathrm{Jy}$, and the minimum detectable flux density for channel-width periodic signal $S_{\min,\mathrm{P}}$ is $45.56 \,\mathrm{Jy}$.

The minimum detectable flux density can be used to estimate the minimum luminosity detection threshold based on the distance of the source $d$, which can be quantified by minimum equivalent isotropic radiated power (EIRP) of the antenna:
\begin{equation}
  \mathrm{EIRP}_{\min}=4\pi d^2 S_{\min}.
  \label{EIRP}
\end{equation}
For the closest target, Barnard's star, the $\mathrm{EIRP}_{\min}$ is $3.98\times10^8 \,\mathrm{W}$ for narrowband signal search, and the $\mathrm{EIRP}_{\min}$ is $1.80\times10^{10} \,\mathrm{W}$ for channel-width periodic signal search.
\subsection{Figures of Merit}\label{subsec:FiguresofMerit}
\begin{figure}[htpb]
  \centering
  \includegraphics[width=0.75\textwidth]{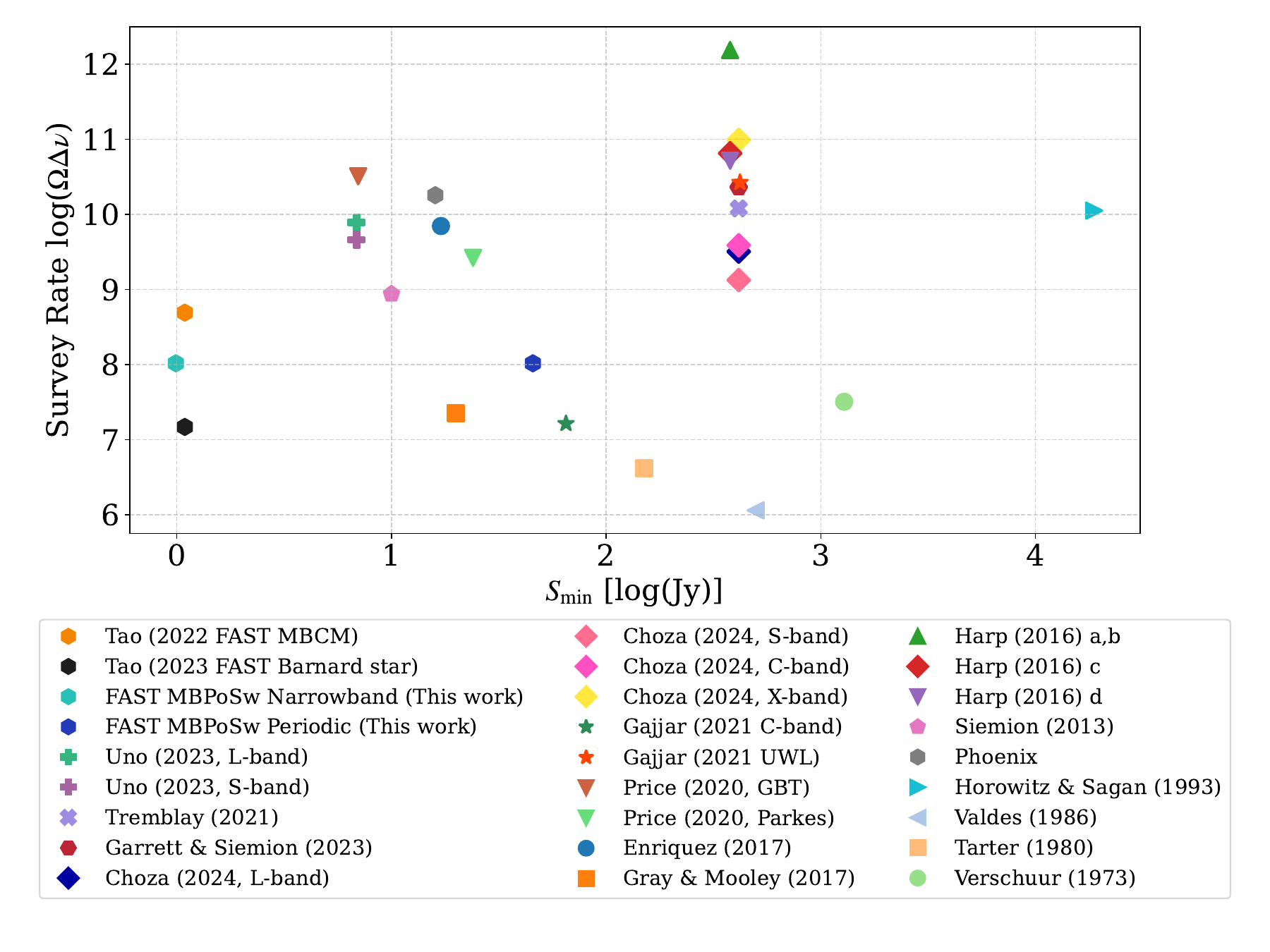}
  \caption{\label{fig:DFM_fig}  $S_{\min}$ versus survey rate $(\Omega\Delta \nu)$ for this work and some previous SETI projects. The plotted points for past surveys are from \cite{1973Icar...19..329V,1980Icar...42..136T,1986Icar...65..152V,1993ApJ...415..218H,2013ApJ...767...94S,2016AJ....152..181H,2017AJ....153..110G,2017ApJ...849..104E,2020AJ....159...86P,2020PASA...37...35T,2021AJ....162...33G,2022AJ....164..160T,2023AJ....166..190T,2023MNRAS.519.4581G,2024AJ....167...10C,2023MNRAS.522.4649U}.}
\end{figure}

\begin{figure}[htpb]
  \centering
  \includegraphics[width=0.75\textwidth]{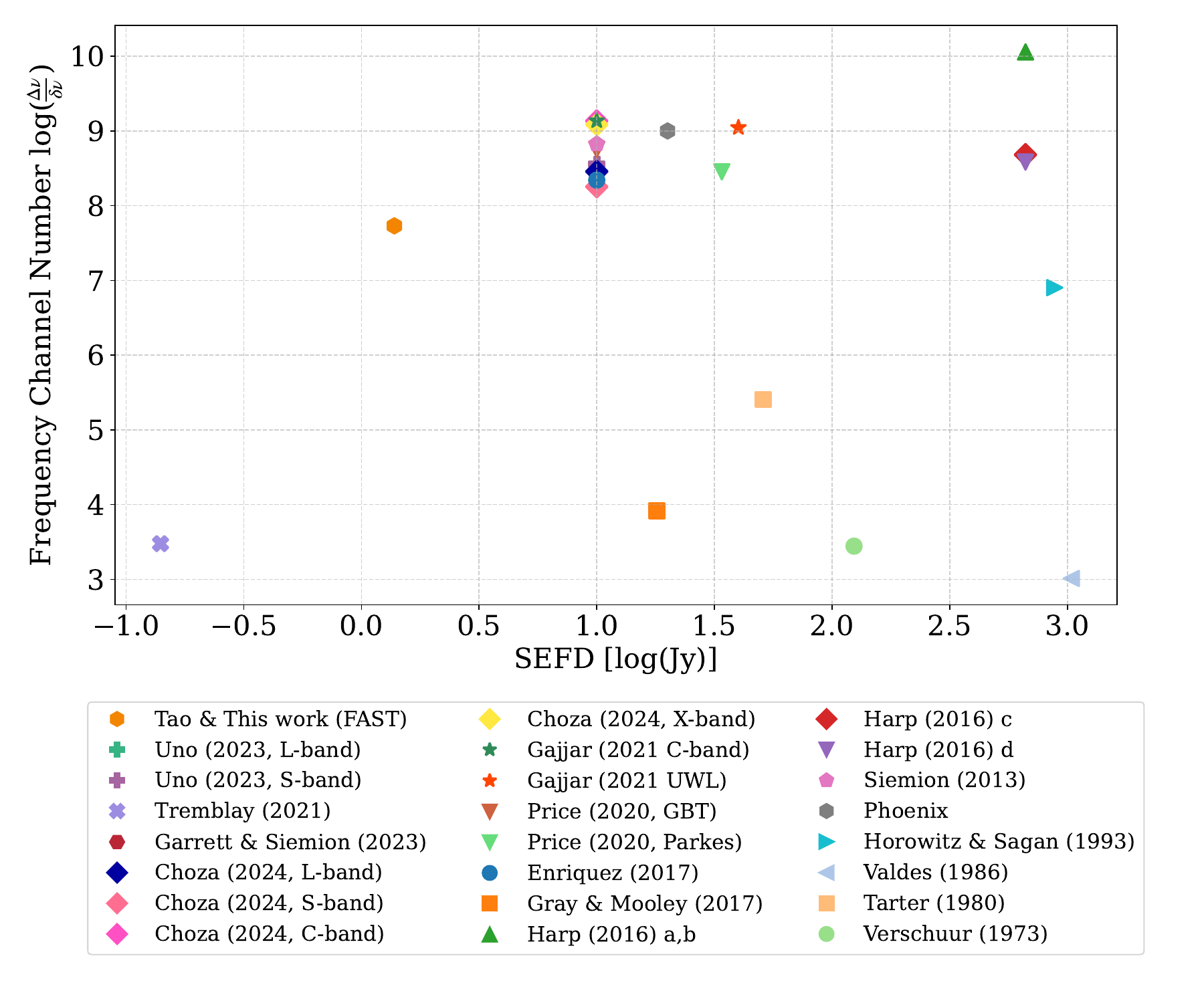}
  \caption{\label{fig:SSFM_fig}  SEFD versus frequency channel number $(\Delta \nu/\delta\nu)$ for this work and some previous SETI projects. The values of the plotted points for past surveys are the same as the references in Figure \ref{fig:DFM_fig}.}
\end{figure}

\begin{figure}[htpb]
  \centering
  \includegraphics[width=0.75\textwidth]{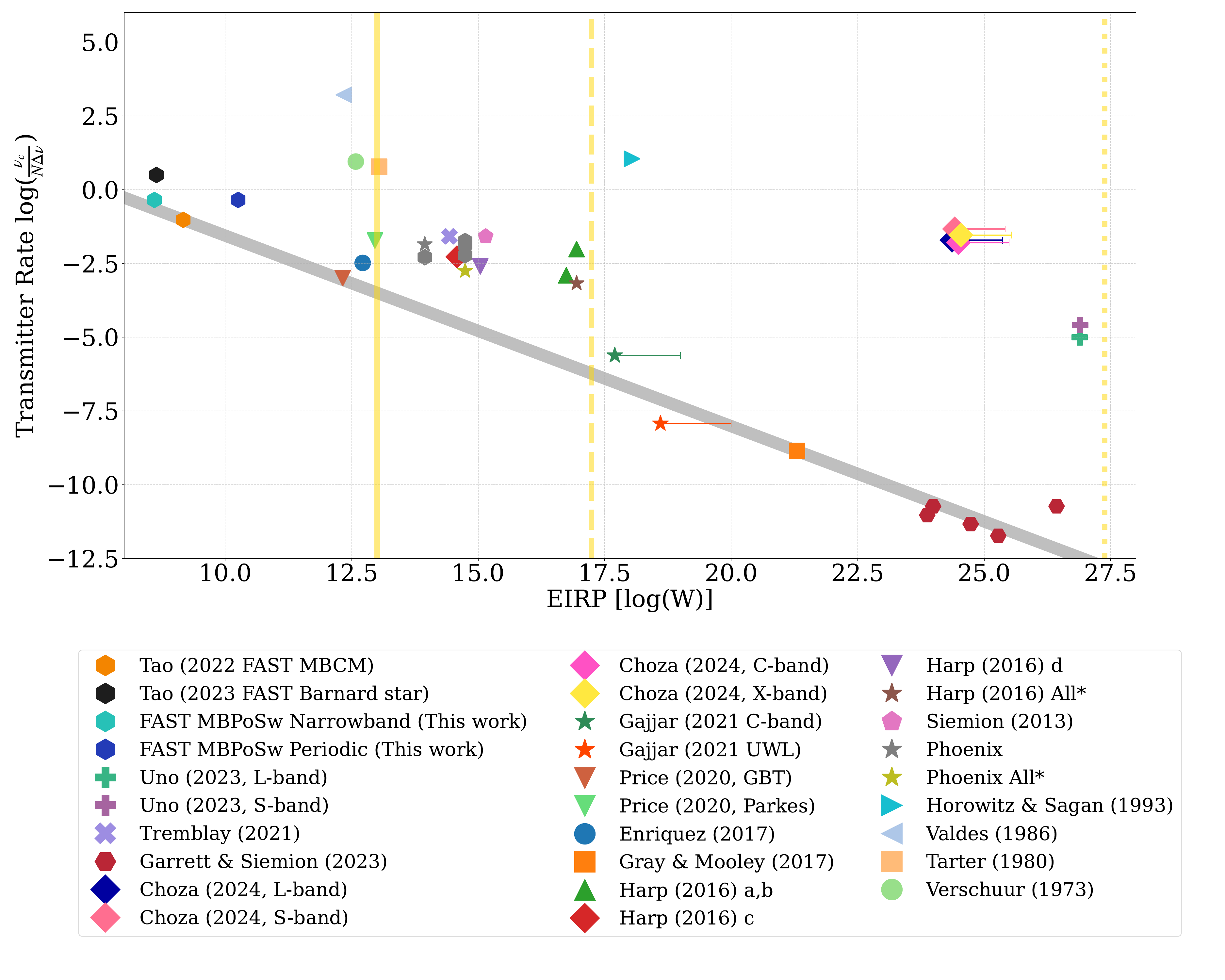}
  \caption{\label{fig:EIRP-TR}  EIRP versus transmitter rate $(N\Delta \nu/\nu_c)^{-1}$ for this work and some previous SETI projects. The solid yellow line indicates the EIRP of the AO planetary radar, the dashed line indicates the typical energy usage of a Kardashev Type I civilization, and the dotted line is the typical energy usage of a Kardashev Type II civilization. The values of the plotted points for past surveys are the same as the references in Figure \ref{fig:DFM_fig}. The gray thick line is a fit between the most constraining data points for the transmitter rate \citep{2017AJ....153..110G}  and $\mathrm{EIRP_{\min}}$ \citep{2022AJ....164..160T}.} 
\end{figure}

\begin{figure}[htpb]
  \centering
  \includegraphics[width=0.75\textwidth]{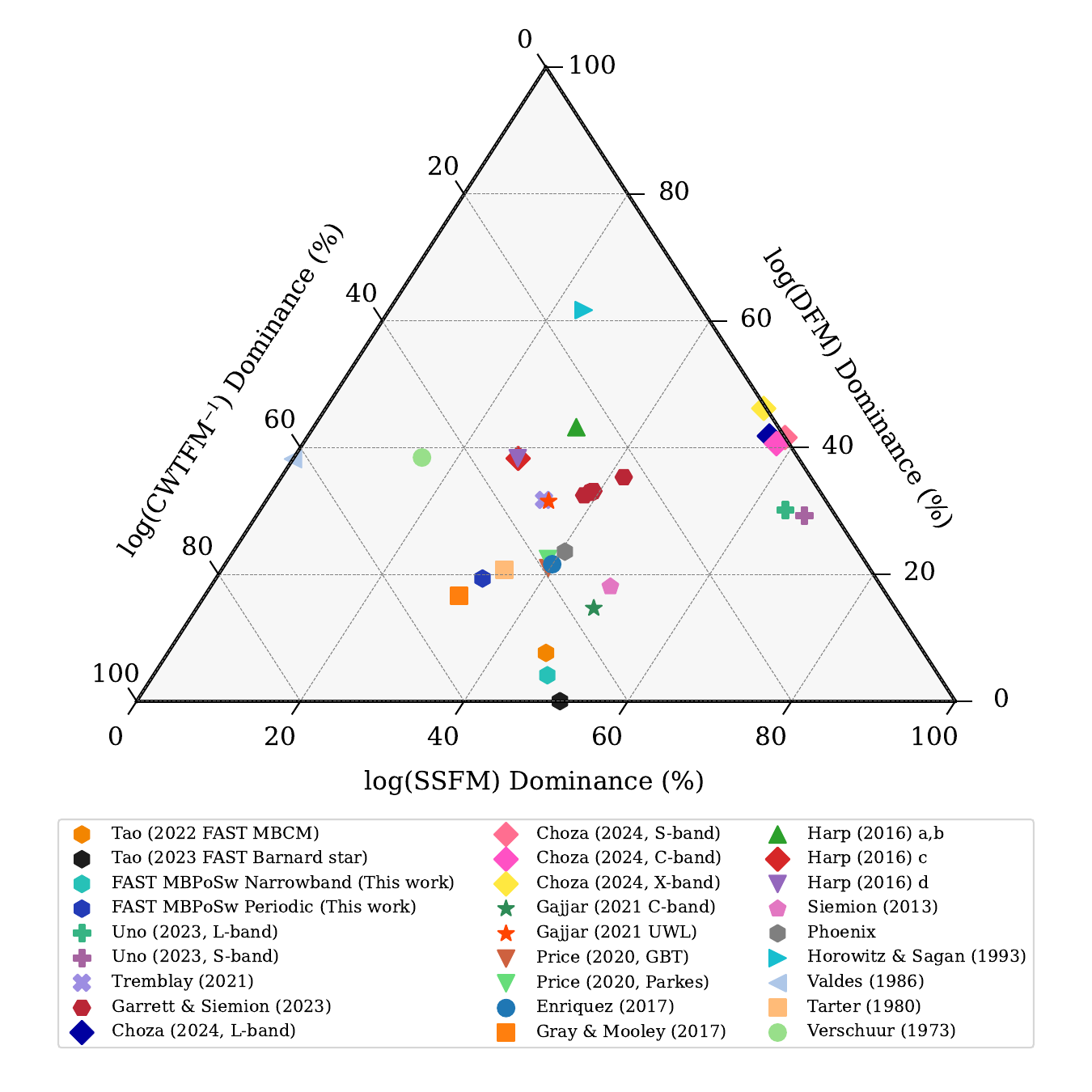}
  \caption{\label{fig:FoMternary} Ternary plot of the normalized FoM dominances for the SETI works. Each FoM dominance is the percentage of its normalized log value divided by the sum of three normalized log values.}
\end{figure}

There are numerous figures of merit (FoM) for characterizing the performance of SETI observations, and one of the most famous FoM is the Drake Figure-of-Merit \citep[DFM]{1984sswg.book.....D} which is commonly defined as
\begin{equation}
  \mathrm{DFM}=\frac{\Omega\Delta \nu}{S_{\min}^{3/2}},
\end{equation}
where $\Omega$ is the total sky coverage. Figure \ref{fig:DFM_fig} shows the survey rate $(\Omega\Delta \nu)$ and sensitivity for different SETI works. Since the FAST observations only focus on several or tens of stars, the sky coverages and survey rate of these works are relatively small.
\cite{2017ApJ...849..104E} introduced a Survey Speed Figure-of-Merit (SSFM) to describe the efficiency of surveys in relation to the telescope and instrumentation used, which can be defined as
\begin{equation}
  \mathrm{SSFM}\propto\frac{\Delta \nu}{\mathrm{SEFD}^2\delta\nu},
\end{equation}
where $\Delta \nu/\delta\nu$ is the frequency channel number. Figure \ref{fig:SSFM_fig} illustrates the relative survey speed comparison in frequency for different SETI works.
For continuous narrowband signal search, a better figures-of-merit, continuous wave transmitter figure of merit (CWTFM) can be applied in targeted observations, which is defined as \citep{2017ApJ...849..104E}
\begin{equation}
  \mathrm{CWTFM}=\zeta_\mathrm{ref} \frac{\mathrm{EIRP}}{N\Delta \nu/\nu_c},
\end{equation}
where $N$ is the number of stars in a given pointing to which one can detect a signal of strength EIRP, $\nu_c$ si the central frequency of the observation and $\zeta_\mathrm{ref}$ is a normalization factor. Figure \ref{fig:EIRP-TR} is the comparison of this work with some historic SETI projects.
The extraordinary sensitivity of FAST guarantees that even the weakest signals it can detect would be readily achievable by contemporary human technology. We also make a comprehansive comparison for these SETI works using normalized DFM, SSFM and CWTFM$^{-1}$ in log scale, which is illustrated in Figure \ref{fig:FoMternary}. 
This ternary plot reveals the trade-offs made between instrumental capabilities  and observational strategies for each SETI works. The FAST observations are positioned near the center of the SSFM axis, indicating that the primary strengths lie in a combination of the survey speed with log(SSFM) dominance $\sim$ 50\% and efficacy of the observational strategy with log(CWTFM$^{-1}$) dominance $\sim$ 50\%, wihch is mainly contributed by unprecedented sensitivity. 

For the periodic signal search, \cite{2023AJ....165..255S} define a periodic spectral signaltransmitter figure of merit (PSSTFM) to quantify the completeness of the observation, wihch can be expressed as
\begin{equation}
    \mathrm{PSSTFM}=\zeta_\mathrm{ref} \frac{\mathrm{EIRP}}{N\Delta \nu/\nu_c\log\left(P_{\max}/P_{\min}\right)\log\left(\delta_{\max}/\delta_{\min}\right)}.
\end{equation}
Following the normalization in \cite{2023AJ....165..255S}, we can calculate that the PSSTFM of our observation is about 0.315, which is slightly lower than the corresponding CWTFM value of 0.321 due to the larger ranges in periods and duty cycles, meaning that the periodic signal search is more complete.

\subsection{Bayesian Limits on the Detection of Technosignatures}
Let us assume that detecting an ETI signal should be an independent and extremely rare event, and each observation towards a star can be treated as an independent Bernoulli trial. Let $f$ be the fraction of planetary systems that have technology to emit radio detectable signs, and $\mathcal{P} $ be the probability of detecting an ETI signal within our observation frequency band and the observation time above the $\mathrm{EIRP_{\min}}$. The conditional probability of finding an ETI signal from the data, also called posterior, can be given by 
\begin{equation}
  p(f,\mathcal{P} | \mathrm{data})=\frac{p(\mathrm{data} | f,\mathcal{P})\pi(f,\mathcal{P})}{p(\mathrm{data})}=\frac{\mathcal{L} (f,\mathcal{P}|\mathrm{data})\pi(f)\pi(\mathcal{P})}{\int{\mathcal{L}(f,\mathcal{P}|\mathrm{data})\pi(f)\pi(\mathcal{P})}\mathrm{d}f\mathrm{d}\mathcal{P}},
  \label{posterior_f_p}
\end{equation}
where $p(\mathrm{data} | f,\mathcal{P})=\mathcal{L}(f,\mathcal{P}|\mathrm{data})$ is the likelihood of zero detections in $N$ trials for the given $f$ and $\mathcal{P}$, $\pi(f,\mathcal{P})$ is the joint prior of the given $f$ and $\mathcal{P}$, wihch can be expressed as $\pi(f,\mathcal{P})=\pi(f)\pi(\mathcal{P})$ based on the independence of $f$ and $\mathcal{P}$, and $p(\mathrm{data})$ is the probability of evidence. The marginal posterior of $f$ should be 
\begin{equation}
  p(f | \mathrm{data})=\frac{\int_{0}^{1}\mathcal{L} (f,\mathcal{P}|\mathrm{data})\pi(f)\pi(\mathcal{P})\mathrm{d}\mathcal{P}}{\int{\mathcal{L}(f,\mathcal{P}|\mathrm{data})\pi(f)\pi(\mathcal{P})}\mathrm{d}f\mathrm{d}\mathcal{P}}.
  \label{marginal_posterior_f}
\end{equation}
The likelihood of zero detection from the independent observations towards $N$ targets can be given by binomial distribution
\begin{equation}
  \mathcal{L}(f,\mathcal{P}|\mathrm{data})=(1-f\mathcal{P})^{N}.
\end{equation}
\subsubsection{Uninformative Prior}\label{subsubsec:UninformativePrior}
To reflect a state of prior ignorance, we assign log-uniform prior for $\pi(f)\propto 1/f$ due to our lack of knowledge about its fundamental scale, and uniform prior $\pi(\mathcal{P})=1$ for $\mathcal{P}$. There are 7 targets were observed, and each observation is regarded as discrete trial, then we can place a limit with 95\% confidence interval that fewer than 5.98\% of the stars we observe have narrowband transmitter above $\mathrm{EIRP_{\min}}$ of $3.98\times10^{8}$ W or periodic transmitter above $\mathrm{EIRP_{\min}}$ of $1.80\times10^{10} \,\mathrm{W}$ (See Appendix \ref{subsec:appendix_UninformativePrior} for analytic calculation).

\subsubsection{Updated Prior}
We involve the observations for 33 targets in \cite{2022AJ....164..160T} to update the belief of the prior. The initial priors for $f$ and $\mathcal{P}$ are still the same as Section \ref{subsubsec:UninformativePrior}, while the posterior obtained from the observations of \cite{2022AJ....164..160T} serves as a new prior of this work, i.e., $\pi(f,\mathcal{P} )=p(f,\mathcal{P} | \mathrm{data_1})$. By updating the prior of $f$ and $\mathcal{P}$ via two targeted observations with FAST, we can place a limit with 95\% confidence interval that fewer than 1.41\% of the stars we observe have narrowband transmitter above an $\mathrm{EIRP_{\min}}$ of $3.98\times10^{8}$ W or periodic transmitter above $\mathrm{EIRP_{\min}}$ of $1.80\times10^{10} \,\mathrm{W}$ (See Appendix \ref{subsec:appendix_UpdatedPrior} for analytic calculation).

\subsection{Multibeam Observation Comparison}\label{subsec:MBComparison}
The multibeam methods can comprehensively utilize the beams of the telescope,which dramatically increase observing efficiency by providing simultaneous, wide-area sky coverage. Although they share the same major reflecting surface, each individual beam path possesses unique instrumental characteristics. Properties such as system temperatures, gains, and polarization leakages can vary distinctly from beam to beam. Some of the signal can arise due to fluctuations or differences of these properties among the beams, which may potentially lead to the misinterpretation of data and false positives. The MBCM tracking observation can guarantees uninterrupted on-target integration for the central beam, it creates a fundamental calibration ambiguity. All beams cannot be calibrated since the central beam can only acquire on-source sample while other beam can obtain off-source sample. A valid calibration for the on-source beam is impossible in this state, as it lacks a corresponding background noise measurement. 

Therefore, to make calibration and validate any candidate signal,  switching the central beam to an empty space during the observation is imperative. An alternative calibration strategy is multibeam point-source scanning, often executed using an MultiBeamOTF mode. In this technique, the telescope slews continuously, allowing the beams to scan across a source in R.A. or decl. direction, reconstructing on-source and off-source from a continuous scan. This scanning method is highly effective for observing calibrators, as it allows for the simultaneous characterization of all beams that pass over the source. However, its primary drawback is that the effective on-source integration time for any single beam is inherently brief, limited to the few moments it takes for the beam to transit across the source. When considering both calibration accuracy and the need to maximize on-source integration time for our science targets, multibeam position switching emerges as the superior choice. This interleaved On-Off approach yields a credible positive sample (on-source beam) and a corresponding negative sample (off-source beam observing empty sky) simultaneously. This duality is critical for efficiently gathering calibrated data and for reliably distinguishing signals of interest from instrumental artifacts or RFI, making it the optimal strategy for our targeted SETI observations.

\subsection{Ancillary Science in the Future}
Considering the activities of the target stars, stellar radio bursts can be also included as an ancillary scientific goal in our SETI observation. Stellar radio bursts are typically characterized by a high degree of circular polarization.  After RFI flagging, polarization calibration, and flux calibration mentioned in Section \ref{sec:DataAnalysis}, we detected no significant stellar radio bursts with these features in our dynamic spectra. Although our target stars are known for their magnetic activities, this does not guarantee they were in an active state during our limited observation window. In the future, we plan to conduct long-term monitoring for some of these sources, which will not only enhance the probability of detecting potential ETI signals from these exoplanetary systems but also increase the likelihood of capturing stellar radio bursts. Some of the physical features, such as pseudosinusoidal frequency drift \citep{2022ApJ...938....1L} and polarization variation \citep{2024AJ....167....8L} with parallactic angle can be applied for the analysis for signals of interest in long-term observation.

Furthermore, the asymmetric On/Off observational strategy employed in this observation, while challenging our visual inspection process with an increased number of RFI, has yielded a valuable legacy dataset. This extensive RFI sample library can be systematically analyzed the physical properties of RFIs and be utilized it for the training set for machine learning in the future.

\section{Conclusion}\label{sec:Conclusion}
We perform the first SETI observation with dual digital backends toward nearby planet-hosting systems using position switching observation mode, with FAST L-band multibeam receiver in the frequency range of 1.05-1.45 GHz. The data was recorded on SETI backend with 7.5 Hz frequency resolution and 10 s sample time, as well as on psr backend with 49.152 $\mu$s sample time and 0.122 MHz frequency resoltion, respectively. We search for narrowband drifting signal with drift rate within diversified drift rate ranges and S/N above 10, as well as periodic pulsed signal with $P\in[0.12\ s,100\ s]$ and $\delta\in[0.1,0.5]$. 

Almost all of the candidates we obtain are obvious RFIs in false positives or in the range of known frequencies of RFI sources. After frequency exclusion and visual inspections, we can place a constrain no solid evidence for any narrowband transmitters with EIRP above $3.98\times10^8 \,\mathrm{W}$ or periodic transmitter with EIRP above $1.80\times10^{10} \,\mathrm{W}$ emitting radio signal within the observation band at the 95\% conﬁdence interval. Stellar radio bursts are neither detected in the observations for these active stars.

In the future, we plan to carry out long-term observation for some of nearby stars, so that we can increase the probability for detecting ETI signals and some useful physical criteria can be applied in the analysis. The probability of detecting stellar radio bursts, which are tightly relavant to the planetary habitabilities around active stars. We also plan to explore the physical properties of RFIs statistically or via machine learning.

\section*{ACKNOWLEDGMENTS}
We sincerely appreciate the referee's excellent feedback and valuable suggestions, which help us greatly refine the analysis process of the narrowband signal search and periodic signal search in our manuscript. We are grateful to the staff of FAST for their help with the arrangements of our observations. We also sincerely thank Yu Hu and Bo-Lun Huang for insightful advice on periodic signal search. This work was supported by National Key R\&D Program of China, No.2024YFA1611804, the Shandong Provincial Natural Science Foundation (ZR2024QA180) and the China Manned Space Program with grant No. CMS-CSST-2025-A01. This work made use of the data from FAST (Five-hundred-meter Aperture Spherical radio Telescope).  FAST is a Chinese national mega-science facility, operated by National Astronomical Observatories, Chinese Academy of Sciences.


\appendix

\section{RFI Flagging}\label{subsec:appendix_RFIFlagging}
The processes of persistent RFI regions flagging in this work are as follows:
\begin{enumerate}
  \item We take the time average of the 2D dynamic spectrum to produce a 1D spectrum $\mathcal{S} (\nu_i)$, and subsequently calculate the median $m_0$ and the MAD $\mathrm{median}(|\mathcal{S} (\nu_i)-m_0|)$ of the averaged spectrum.
  \item Strong RFI regions are preliminarily identified by $\mathcal{S} (\nu_i)>m_0+5\cdot\mathrm{MAD}$ with their corresponding region width $w_k$. The characteristic width of RFI region can be determined by $W_c=\mathrm{median}\{w_1, w_2, \dots, w_k\}$.
  \item The MAD filter is applied to the rest of the unflagged averaged spectrum again to obtain a set of relatively quiescent channels $\mathcal{C}_q $, which are used to produced a continuous interpolated spectrum $\mathcal{S}_\mathrm{interp}(\nu_j)$.
  \item The window size for baseline smoothing, set to be significantly larger than this characteristic width, can be adaptively determined by $W_\mathrm{smo}=\mathrm{odd}(\lfloor 5W_c\rfloor) $ to ensure the baseline estimation is not biased by the RFI morphology. The smooth background baseline is then derived by applying a median filter of this width to the interpolated spectrum: $\mathcal{B} (\nu_i)=\mathrm{median}\{\mathcal{S}_\mathrm{interp}| j\in [i-w,i+w]\}$,where $w=(W_\mathrm{smo}-1)/2$.
  \item The residuals can be obtained from the quiescent channels by $\mathcal{R} (\nu_i)=\mathcal{S}(\nu_i) -\mathcal{B}(\nu_i) $, with the corresponding robust standard deviation of these residuals $\sigma_\mathrm{res}=1.4826\cdot\mathrm{median}\{|\mathcal{R} (\nu_i)|, j\in \mathcal{C}_q \}$. Any frequency channel with residual larger than $3\sigma_\mathrm{res}$ threshold are flagged as the weak RFI.
\end{enumerate}
In all observations, frequency channels with severe RFI contamination predominantly occurred within the range of approximately 1140 MHz to 1290 MHz, where RFI levels are extremely elevated compared to normal level. Additionally, weak RFI could also appear in over a dozen frequency channels around approximately 1090 MHz in some observations. This result is consistent with the RFI monitoring reports of FAST\footnote{\url{https://fast.bao.ac.cn/cms/category/rfi_monitoring_en}}.

\section{Calculations for maximum drift rate}\label{subsec:appendix_MDR}
The total drift rate caused by the rotations and orbits of Earth and the exoplanet, which can be calculated by the Equation (30) in \cite{2022ApJ...938....1L}, and the corresponding MDR are listed in Table \ref{tab:exoplanet_mdr}. The initial frequency $\nu_0$ of the signal is 1.45 GHz. Based on the theoretical MDR calculated for each exoplanet, we adopt different MDRs for the targets. For each target, if no exoplanet shows a theoretical MDR greater than the TurboSETI default of 4 $\mathrm{Hz \ s^{-1}}$, the MDR used in the search is set to be 4 $\mathrm{Hz \ s^{-1}}$. While if there are exoplanets with theoretical MDR greater than 4 $\mathrm{Hz \ s^{-1}}$, the MDR adopted in the search should be set to be an integer immediately above the target's maximum theoretical MDR. Such choice ensures coverage of the full physically expected drift rates while keeping the search space minimal.
\begin{deluxetable}{llcc}[htpb]
\tablecaption{Theoretical MDR values for exoplanets and used values for each target.\label{tab:exoplanet_mdr}}
\tablehead{
\colhead{Exoplanet} & \colhead{Host star} & \colhead{MDR ($\mathrm{Hz \ s^{-1}}$)} & \colhead{MDR ($\mathrm{Hz \ s^{-1}}$) in the search}
}
\startdata
Barnard's Star b & \multirow{4}{*}{Barnard's Star} & 8.949048  & \multirow{4}{*}{14} \\
Barnard's Star c &                                 & 7.451418  &                      \\
Barnard's Star d &                                 & 13.509880 &                      \\
Barnard's Star e &                                 & 3.633798  &                      \\
Ross 128 b       & Ross 128                        & 3.449406  & 4                    \\
Gliese 581 b     & \multirow{3}{*}{Gliese 581}     & 3.909113  & \multirow{3}{*}{8}   \\
Gliese 581 c     &                                 & 1.429715  &                      \\
Gliese 581 e     &                                 & 7.441178  &                      \\
Upsilon Andromedae A b & \multirow{3}{*}{Upsilon Andromedae A} & 9.724095 & \multirow{3}{*}{10} \\
Upsilon Andromedae A c &                            & 0.263266  &                      \\
Upsilon Andromedae A d &                            & 0.190067  &                      \\
55 Cancri A b    & \multirow{5}{*}{55 Cancri A}    & 2.070588  & \multirow{5}{*}{14}  \\
55 Cancri A c    &                                 & 0.636947  &                      \\
55 Cancri A d    &                                 & 0.180898  &                      \\
55 Cancri A e    &                                 & 13.574282 &                      \\
55 Cancri A f    &                                 & 0.225611  &                      \\
Lalande 21185 b  & \multirow{2}{*}{Lalande 21185}  & 2.230676  & \multirow{2}{*}{4}   \\
Lalande 21185 c  &                                 & 0.181605  &                      \\
Wolf 359 b       & Wolf 359                        & 0.180902  & 4                    \\
\enddata
\end{deluxetable}

\section{Analysis for the Periodic Pulsed Signal Search}
\subsection{Statistical Attributions for the Candidate Number Differences}\label{sec:appendix_cand_stat}

Given that the candidate number differences of the observations in 2024 November 15 (hereafter $F_1$ 241115) and 2025 July 9 (hereafter $F_1$ 250709) are in two orders of magnitude, we make statistical comparison and test for the candidates of the observations in these two days. The candidates of the 30-second series FFA search are not included in $F_1$ 250709, since we only carry out 6-minute FFA search for the observation in 2024 November 15. Figure illustrates the statistical distributions and the violin plots for the periods, frequencies and the signal-to-noise ratios of these candidates. Compared with $F_1$ 241115, $F_1$ 250709 exhibits periods concentrated at shorter values with several narrow modes instead of broad distribution with a long tail to large periods, radio-frequency occupancy spanning larger band range with clear multimodalities in specific RFI regions rather than the narrower band around 1.18-1.26 GHz, and the S/N distribution with a much heavier upper tail reaching several hundred. Taken together, these patterns indicate substantial distributional differences for the features of these two groups of candidates.
\begin{figure}[htpb]
  \centering
  \includegraphics[width=\textwidth]{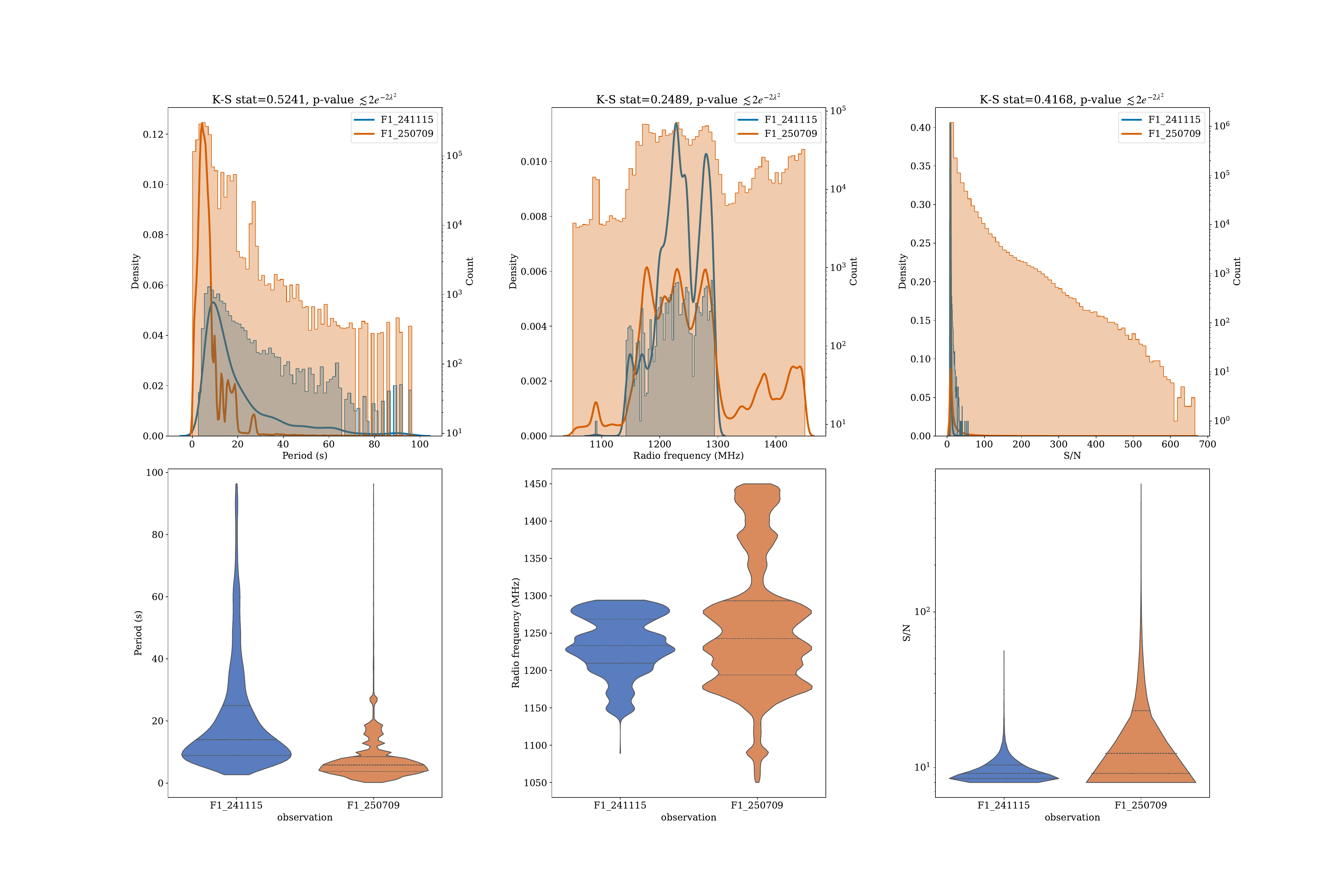}
  \caption{\label{fig:blipss_season_statistics} Statistics for the periods (first column), frequencies (second column) and the signal-to-noise ratios (third column) of these candidates of the observations in 2024 November 15 (blue) and 2025 July 9 (orange). The subfigures in the top row are the statistical distributions while the subfigures in the bottom row is the violin plots.}
\end{figure} 
We also carry out two-sample Kolmogorov-Smirnov test \citep{an1933sulla,smirnov1939estimation,massey1952distribution,miller1956table} for these two groups of periods, frequencies and the signal-to-noise ratios. The null hypothesis is that the two samples come from a specified distribution. For $k$ independent and identically distributed ordered observations $X_i$, the empirical distribution function $F_k$ is
\begin{equation}
  F_{k}(x)=\frac{1}{k}\sum_{i = 1}^{k}  1_{(-\infty ,x]}(X_{i}),
\end{equation}
where $1_{(-\infty ,x]}(X_{i})$ is the indicator function, equal to 1 if $ X_{i}\leq x$ and equal to 0 otherwise. The Kolmogorov-Smirnov statistic is given by
\begin{equation}
  D=\sup_x |F_{1,n}(x)-F_{2,m}|,  
\end{equation}
where $\sup$ stands for the largest value over all possible values of $x$. For large sample, the null hypothesis is rejected at $\alpha \approx 2e^{-2\lambda^2}$, where $\lambda$ is given by \citep{Aryeh1956asymptotic,Pascal1990the,WEI2012636}
\begin{equation}
  \lambda = D\sqrt{\frac{mn}{m+n}}.
\end{equation}
Based on the p-values of these Kolmogorov-Smirnov tests, we can reject the null hypothesis, indicating that these two samples come from different distribution. These may serve as the attribution that the candidate number differences could be caused by the diverse RFI environment in these two days. The statistic results listed in Table \ref{tab:blipss_candidates} include the events and candidates for both 6-minute and 30-second FFA search of the observation in 2025 July 9, while the the events and candidates of the observation in 2024 November 15 are only searched for 6-minute. Therefore, the different off-source time setting in 2025 July 9 also contributes to the event and candidate number differences. Table \ref{tab:blipss_candidates_6min_and_30sec} lists the statistics of the event and candidate for 6-minute and 30-second FFA saerch of the observation in 2025 July 9.
\begin{deluxetable}{lcccccc}[htpb]
\tablecaption{Candidate and event Statistics for the 6-minute and 30-second FFA saerch from the observation in 2025 July 9.}\label{tab:blipss_candidates_6min_and_30sec}
\tablehead{
\colhead{Source} & \colhead{$F_1$ (6-minute)} & \colhead{$F_2$ (6-minute)} &
\colhead{$F_{1}$ (30-second)} & \colhead{$F_{2}$ (30-second)} &
\colhead{$N_{\mathrm{event}}$ (6-minute)} & \colhead{$N_{\mathrm{event}}$ (30-second)}
}
\startdata
55 Cancri A            & 877078 & 255670 & 21452    & 20843    & 5016877 & 138139 \\
Gliese 581             & 343840 & 123924 & 7392    & 6537    & 5149999 & 61504 \\
Lalande 21185          & 214545 & 78338  & 4255  & 3453  & 960507  & 26748 \\
Upsilon Andromedae A   & 198765 & 62695  & 8163    & 7266    & 1261835 & 67597 \\
Wolf 359               & 182063 & 64551  & 10070 & 7785 & 1130427 & 55678 \\
\hline
Total                  & 1816291 & 585178 & 51332 & 45884 & 13519645 & 349666\\
\enddata
\end{deluxetable}

Since the FFA search for the observation in 2024 November 15 is only carried out in the central beam, we also make a test using the central beam (on-source) and reference beam (off-source, here we select beam 14 as reference beam) for the FFA search. Table \ref{tab:blipss_candidates_comparison} lists the comparison of event and candidate numbers for the search results. The event and $F_1$ candidate numbers of two-beam search are about 1.45 times and 1.66 times that of the central-beam search, and there are additional $F_2$ candidates for the two-beam search. 
\begin{deluxetable}{lcccccc}[htpb]
\tablecaption{Candidate Statistical comparison for Barnard's star and Ross 128 using central beam and two beams in the FFA search.}\label{tab:blipss_candidates_comparison}
\tablehead{
\colhead{Source} & \colhead{$F_1$} & \colhead{$F_2$} &
\colhead{$F_{1}$ (two beams)} & \colhead{$F_{2}$ (two beams)} &
\colhead{$N_{\mathrm{event}}$} & \colhead{$N_{\mathrm{event}}$ (two beams)}
}
\startdata
Barnard's Star         & 6283   & 0      & 11648    & 6022    & 55135 & 82782 \\
Ross 128               & 7603   & 0      & 11439    & 6317    & 59777 & 84932 \\
\hline
Total                  & 13886 & 0 & 23087 & 12339 & 114912 & 167714\\
\enddata
\end{deluxetable}

\begin{figure}[htpb]
  \centering
  \includegraphics[width=\textwidth]{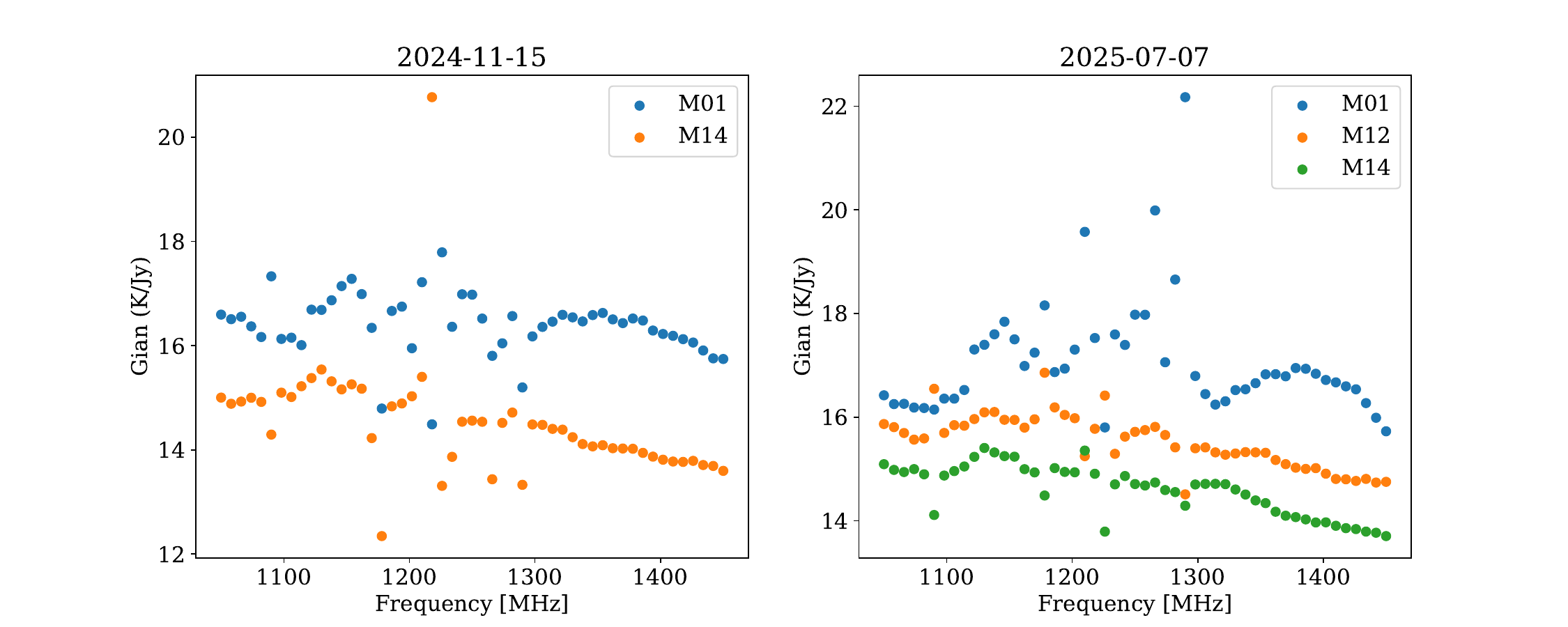}
  \caption{\label{fig:fast-gain-difference} Left panel: The gain differences of beam 1 and beam 14 of FAST in 2024 November 15. Right panel: The gain differences of beam 1, beam 12 and beam 14 of FAST in 2025 July 7 (nearest date to 2025 July 9).}
\end{figure}

We also select the antenna gain measurement data\footnote{\url{https://fast.bao.ac.cn/cms/category/1977569564215562242}} from date closest to the observation date to check the gain responses for the beams used in the search, which are illustrated in Figure \ref{fig:fast-gain-difference}. Generally, at most frequencies, the center beam has a higher gain than the reference beams. Taken the number differences for events and candidates, as well as the gain differences for the beams, we extrapolate that the candidate number differences could also be caused by the diverse gain responses in different beams.

\subsection{Phase-resolved Spectra Analysis}\label{sec:appendix_PRS_analysis}

\begin{figure}[htpb]
\centering
\gridline{
  \fig{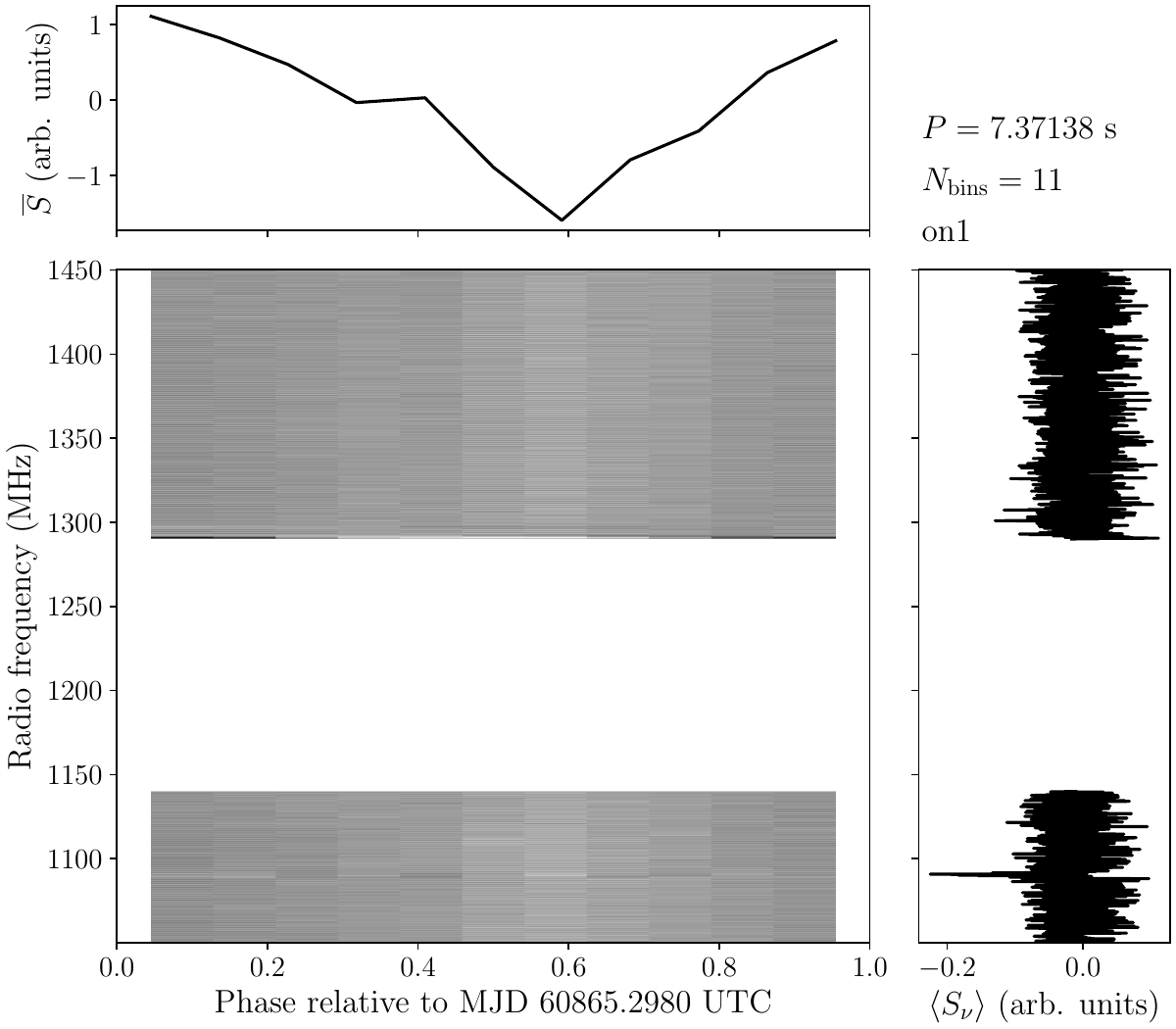}{0.4\textwidth}{}
  \fig{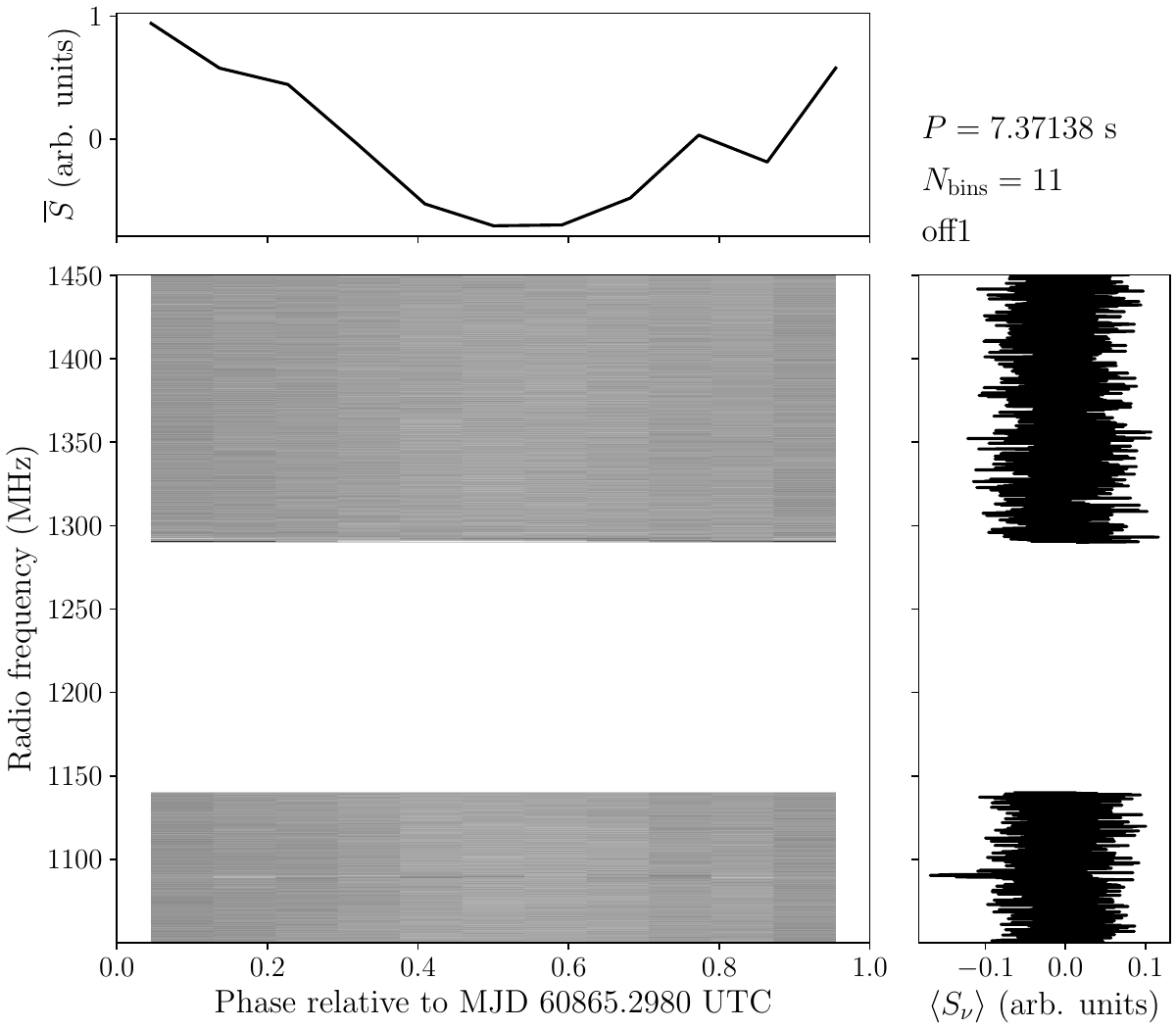}{0.4\textwidth}{}
}\vspace{-6pt}
\gridline{
  \fig{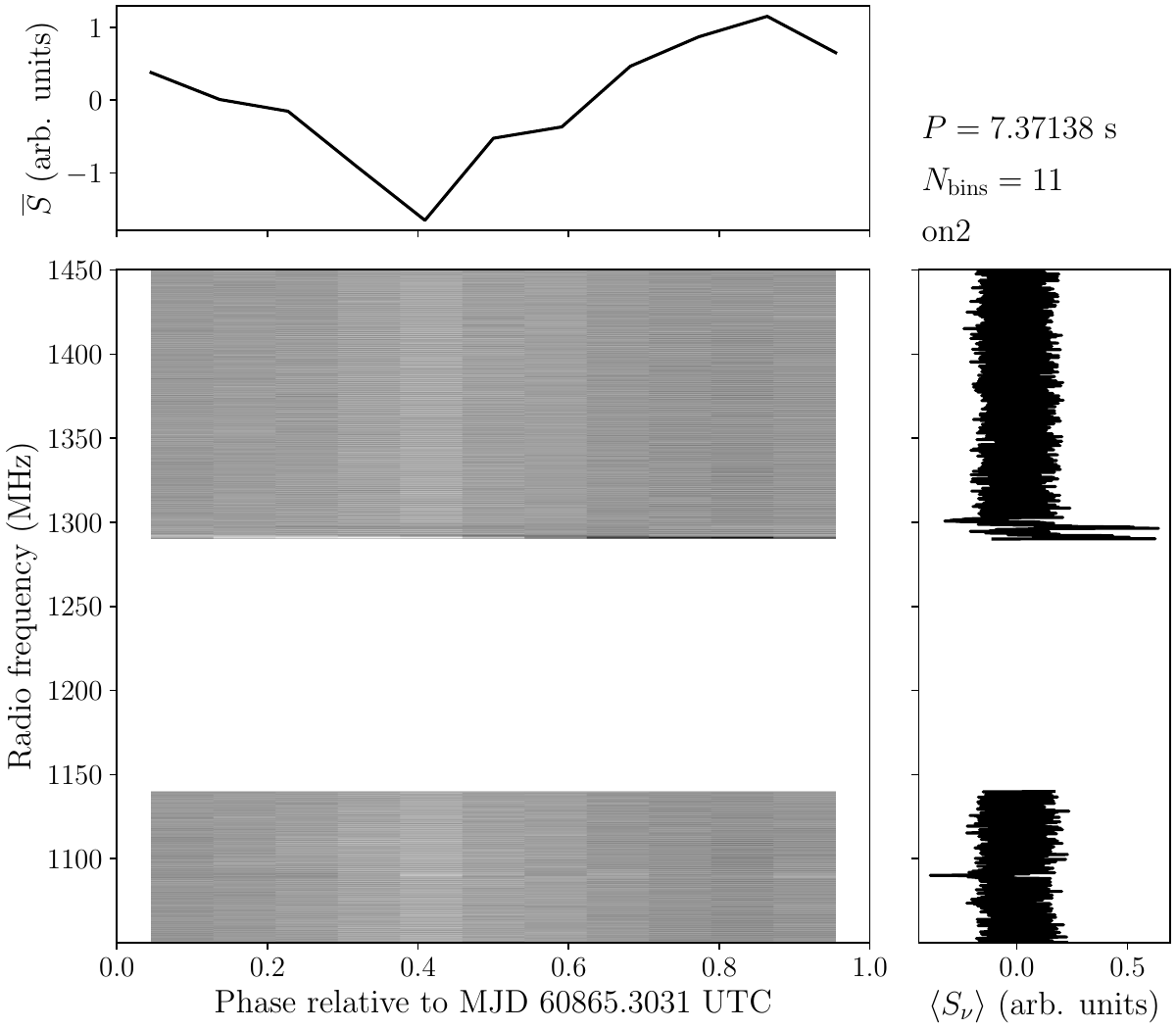}{0.4\textwidth}{}
  \fig{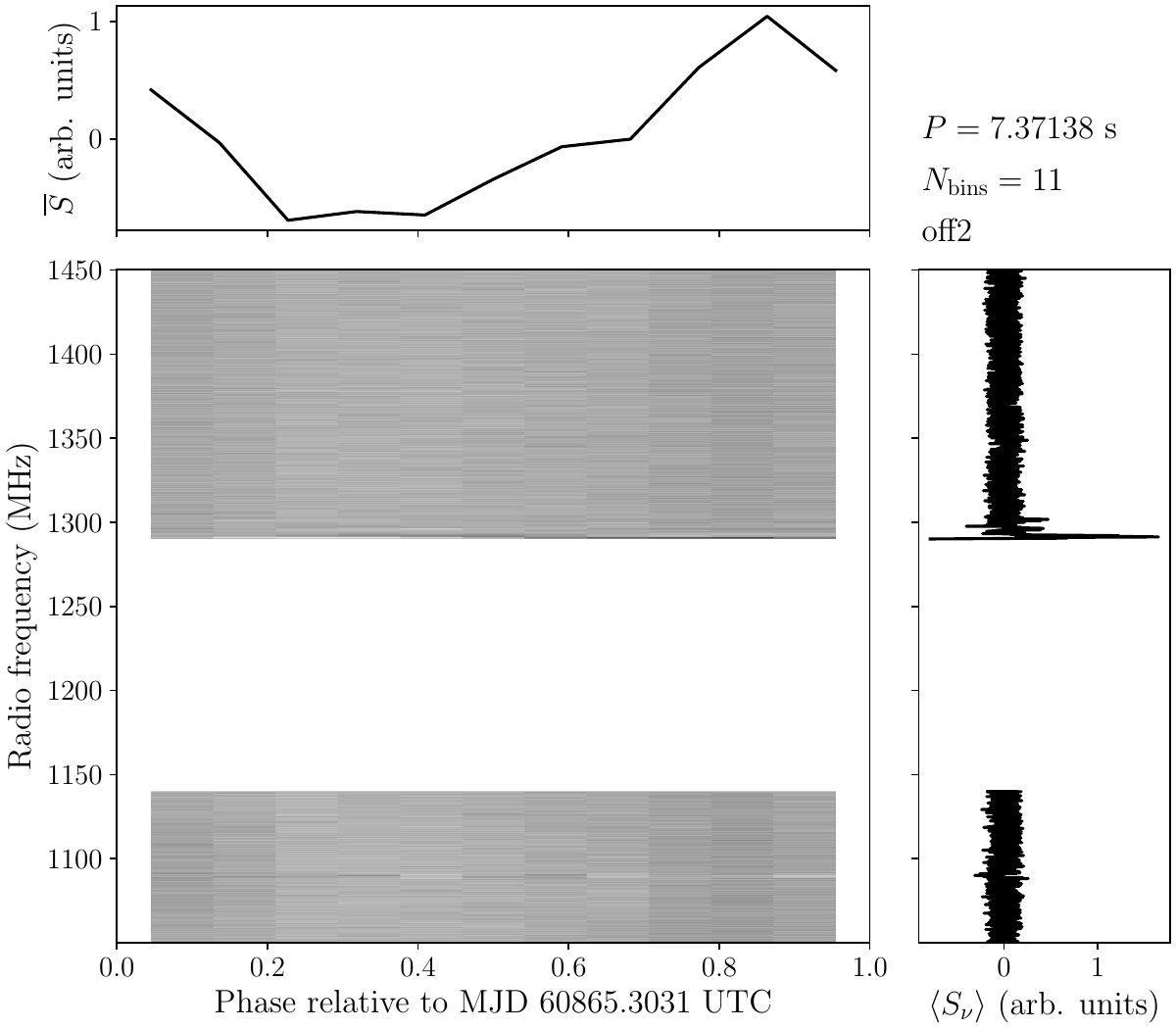}{0.4\textwidth}{}
}\vspace{-6pt}
\gridline{
  \fig{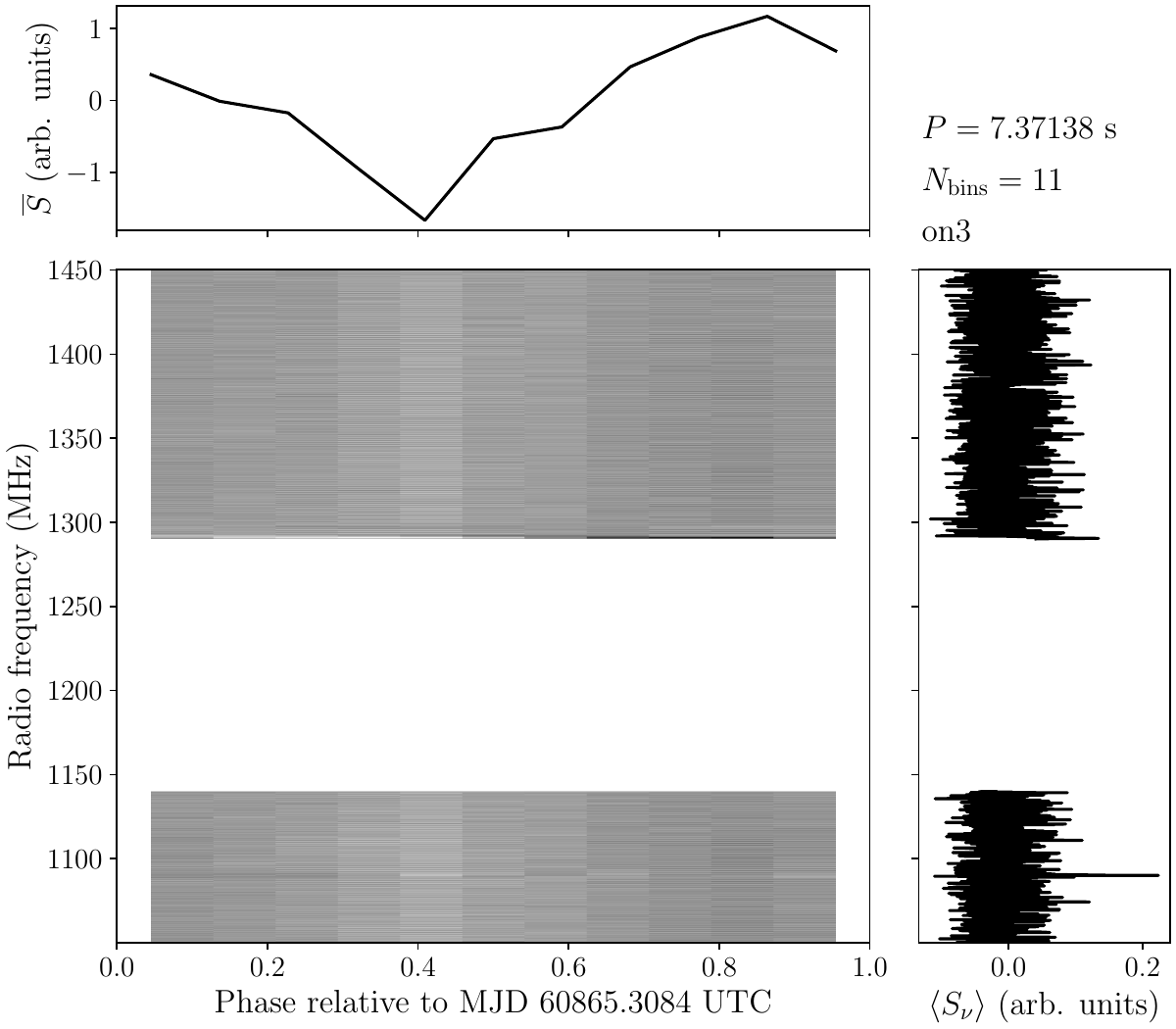}{0.4\textwidth}{}
  \fig{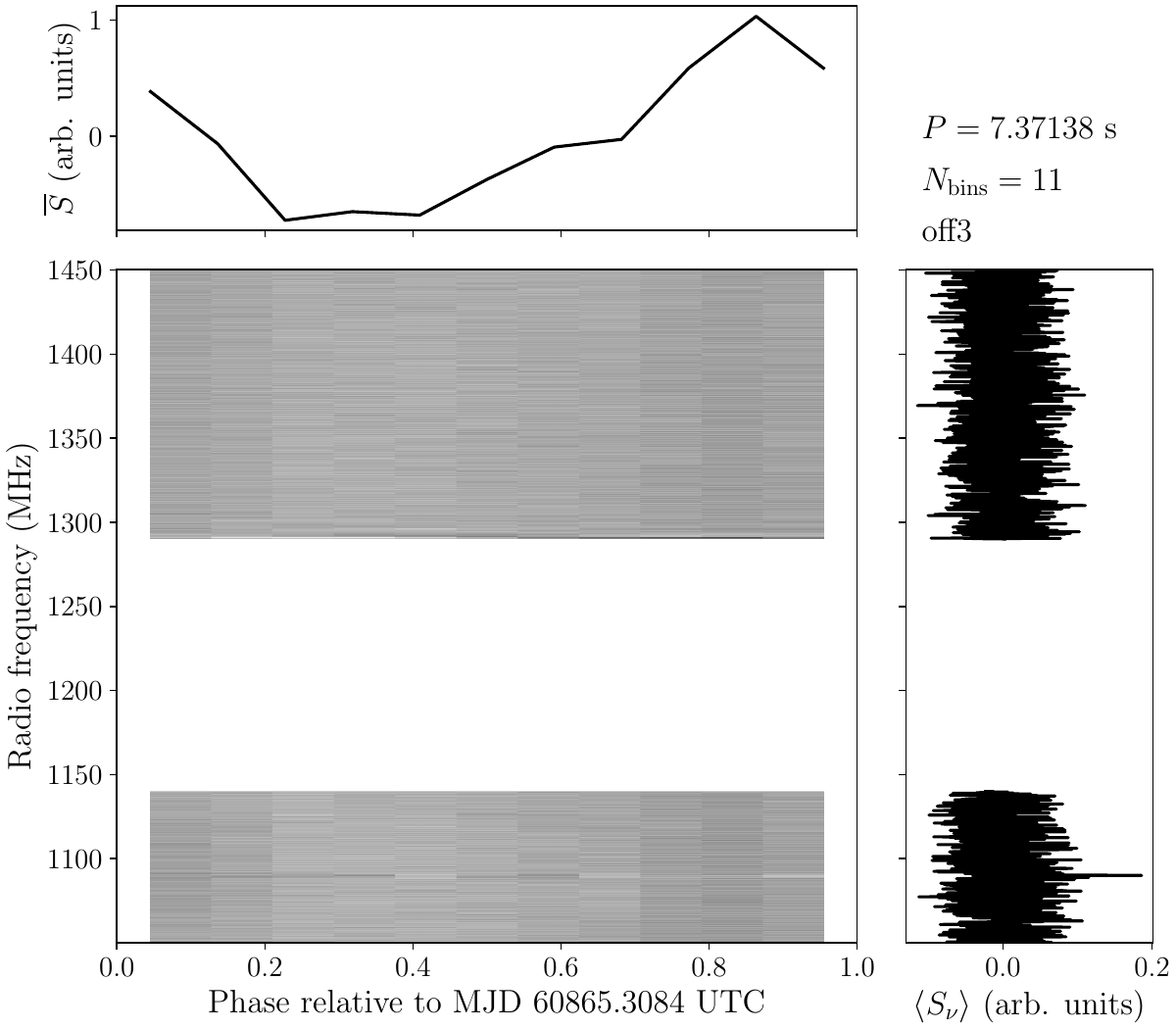}{0.4\textwidth}{}
}
\caption{Phase-resolved spectrum of Lalande 21185 at $P=7.37138 \ \mathrm{s}$ for the three on-source (left column) and off-source (right column) scans. For each panel, top left subplot is the frequency-averaged profile, bottom left subplot is the phase-resolved spectrum and bottom right subplot is the profile spectrum.}
\label{fig:phaseresolved_Lalande21185}
\end{figure}

\begin{figure}[htpb]
\centering
\gridline{
  \fig{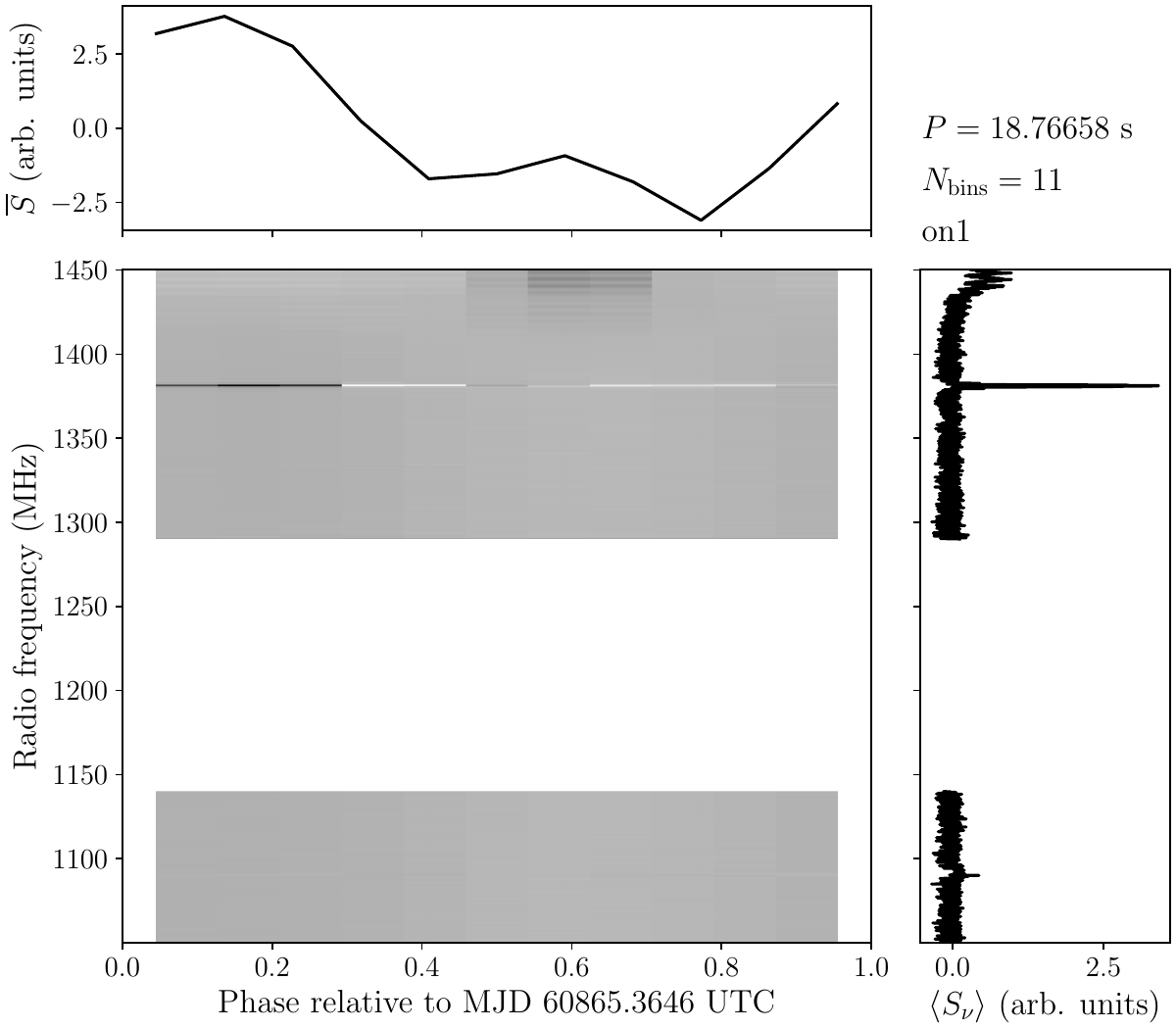}{0.4\textwidth}{}
  \fig{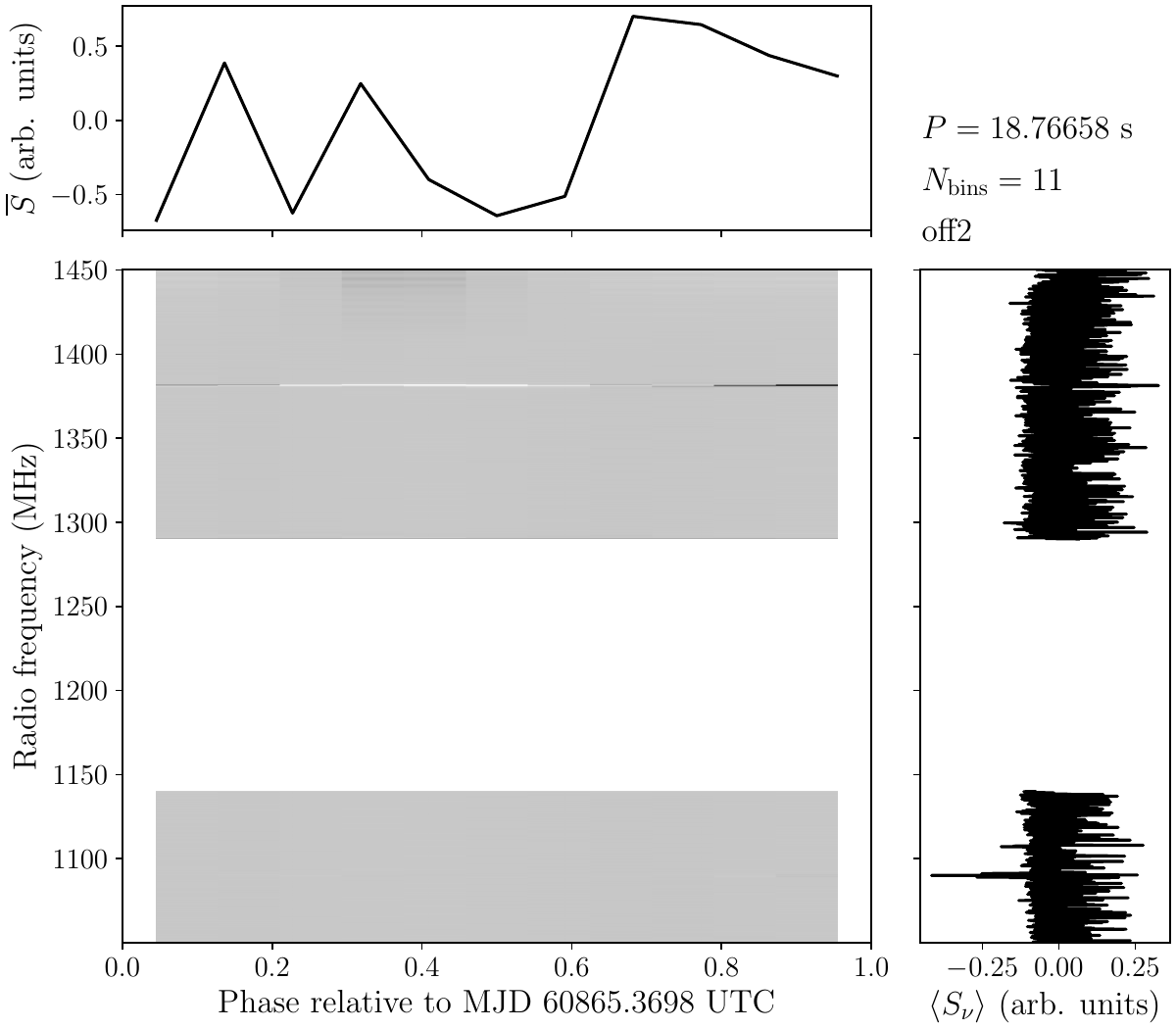}{0.4\textwidth}{}
}\vspace{-6pt}
\gridline{
  \fig{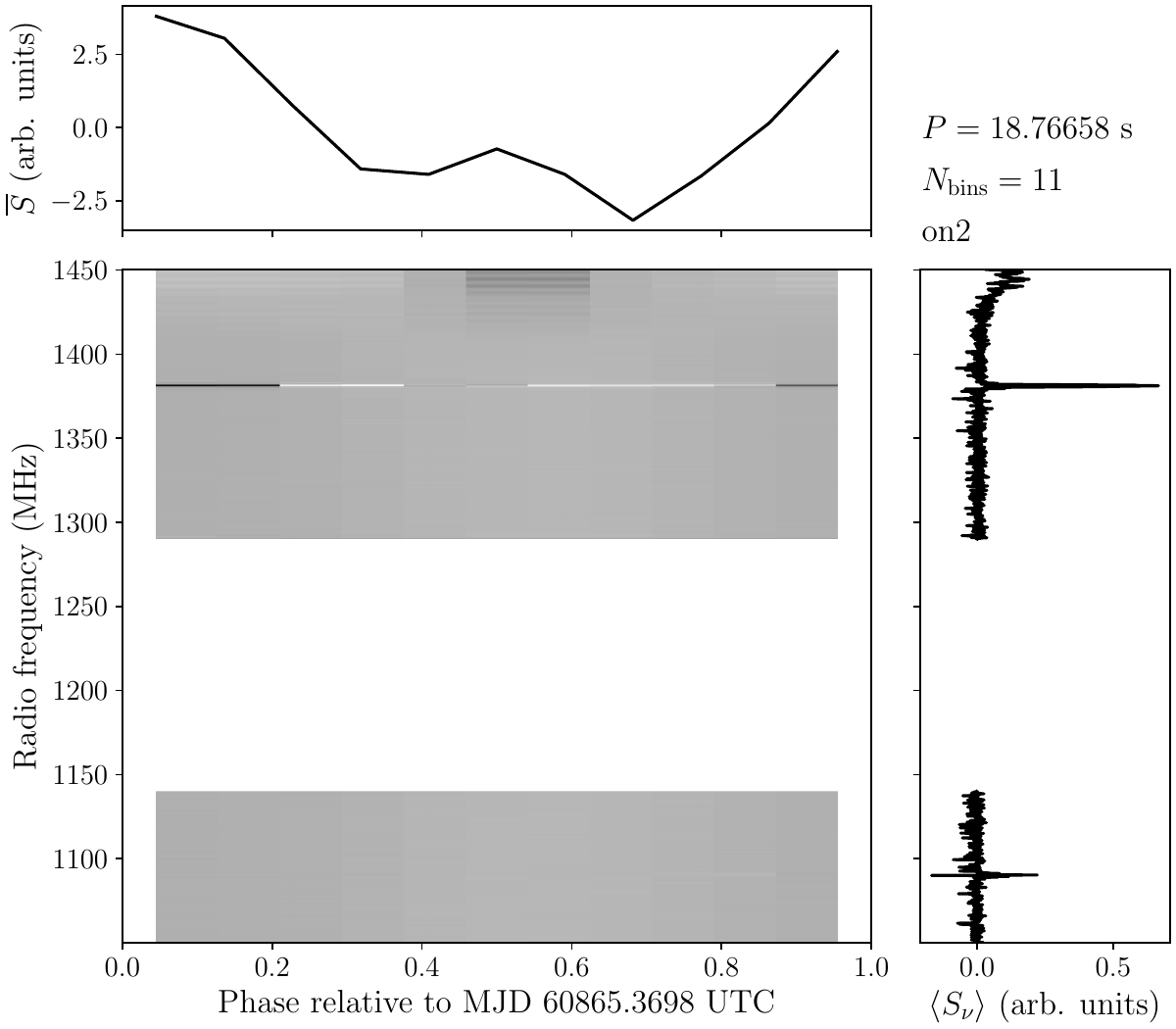}{0.4\textwidth}{}
  \fig{Wolf359_M12_off2_period18.76658.pdf}{0.4\textwidth}{}
}\vspace{-6pt}
\gridline{
  \fig{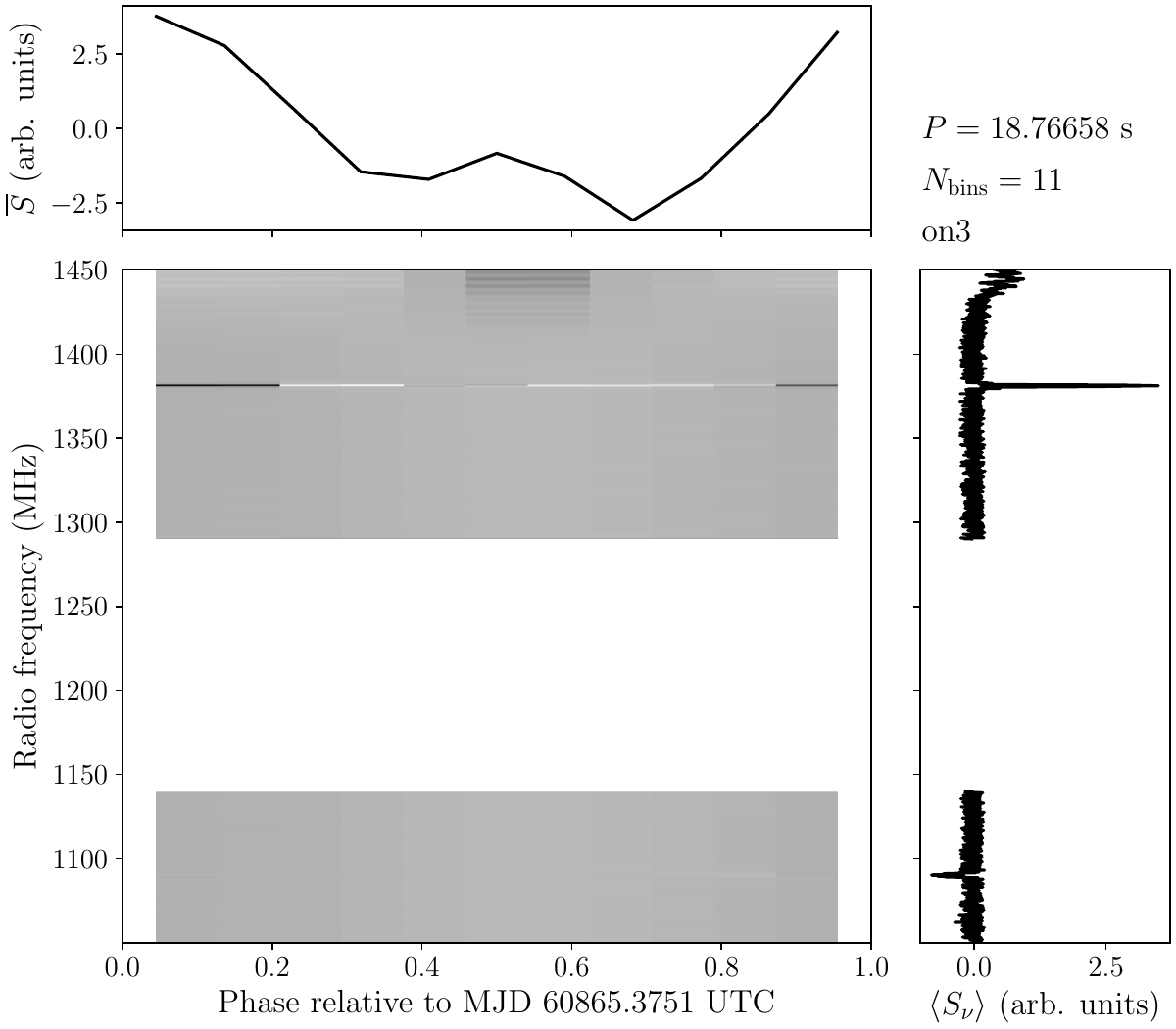}{0.4\textwidth}{}
  \fig{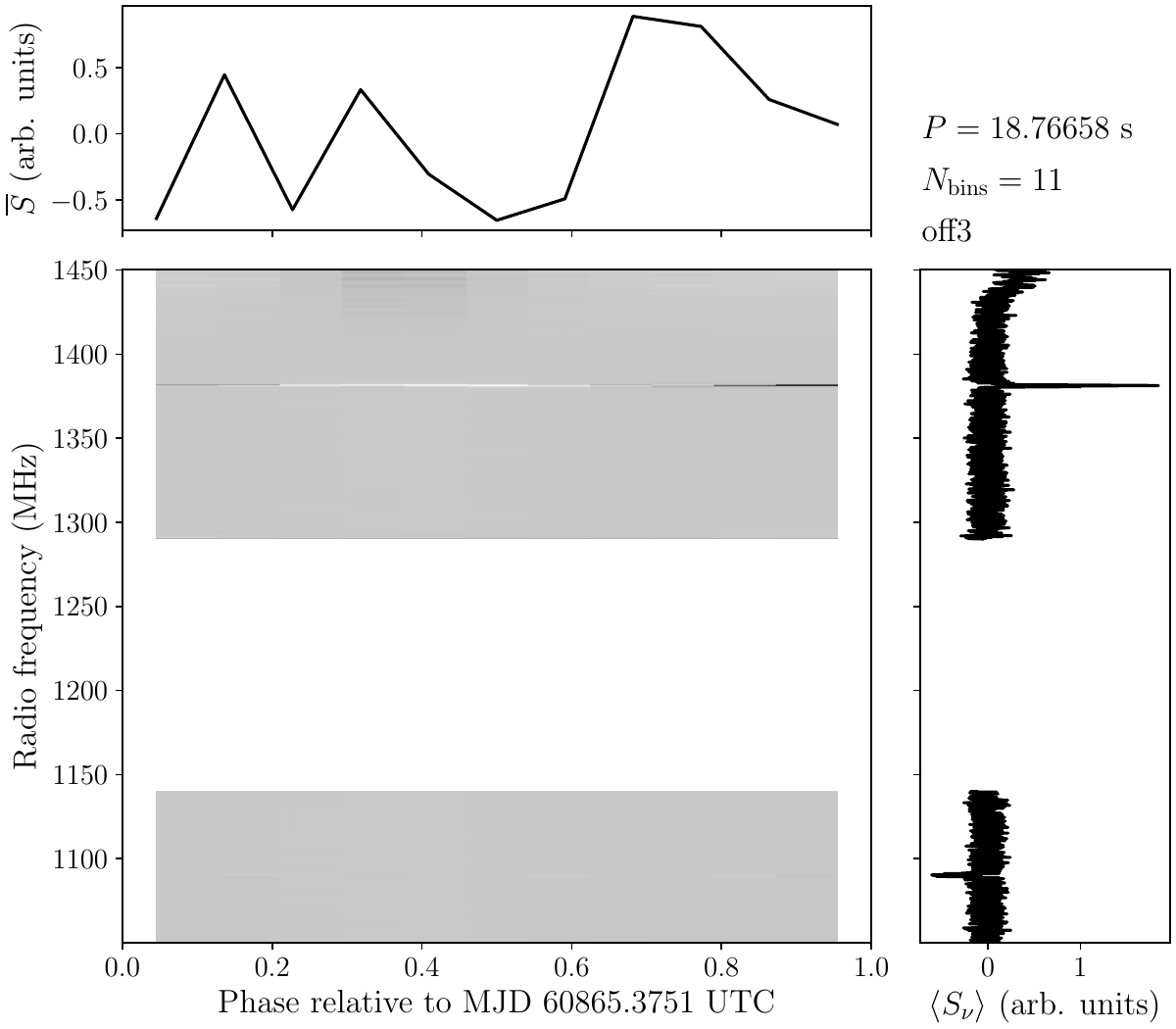}{0.4\textwidth}{}
}
\caption{Phase-resolved spectrum of Wolf 359 at $P=18.76658 \ \mathrm{s}$ for the three on-source (left column) and off-source (right column) scans. The layout of each panel is the same as Figure \ref{fig:phaseresolved_Lalande21185}.}
\label{fig:phaseresolved_Wolf359}
\end{figure}

Since most of the remaining candidates are contributed by the observations of Lalande 21185 and Wolf 359, and these candidates span the L-band at some specific periods, we generate the phase-resolved spectra for these two targets using the periods determined in Figure by the number probability density profile. Figure \ref{fig:phaseresolved_Lalande21185} and \ref{fig:phaseresolved_Wolf359} are the phase-resolved spectra within 1050 MHz - 1140 MHz and 1290 MHz - 1450 MHz of Lalande 21185 at $P=7.37138 \ \mathrm{s}$ and Wolf 359 at $P_\mathrm{main}=18.76658 \ \mathrm{s}$, respectively. Although we can see significant peaks in the profile spectrum Lalande 21185 at about 1290 MHz and Wolf 359 at about 1390 MHz, as well as remarkable phase differences in the phase-resolved spectra of Wolf 359, these phenomena also exist in the off-source scans, we tend to attribute these to periodic fluctuations in noise within  specific frequency channels. The phase-resolved spectra of Wolf 359 at other three secondary peak also reveal no significant broadband pulse. 

\section{Calculation for Bayesian Limits}
\subsection{Uninformative Prior Case}\label{subsec:appendix_UninformativePrior}
The probability of the evidence can be calculated by 
\begin{eqnarray}
  p(\mathrm{data})&&\propto\iint_{0}^{1}{(1-f\mathcal{P})^{N} f^{-1} }dfd\mathcal{P},\nonumber\\
  &&=\int_{0}^{1}{\frac{1-(1-f)^{N+1}}{f^2(N+1)}\mathrm{d}f},\nonumber\\
  &&=\int_{0}^{1}{\sum_{n = 1}^{N+1}\binom{N+1}{n} \frac{(-1)^{n+1}f^{n-2}}{N+1}  \mathrm{d}f},\nonumber\\
  &&= \int_{0}^{1}f^{n-2}\mathrm{d}f +\sum_{n = 1}^{N+1}\binom{N+1}{n}\frac{(-1)^{n+1}}{N+1}  \mathrm{d}f.
  \label{pdata}
\end{eqnarray}
Then, the 95\%  credible upper limit can be obtained by solving
\begin{equation}
  \int_{0}^{f_{\mathrm{upper}}}\frac{p(f | \mathrm{data})}{p(\mathrm{data})}\mathrm{d}f=\int_{0}^{f_{\mathrm{upper}}}\frac{1-(1-f)^{N+1}}{p(\mathrm{data})f^2(N+1)}\mathrm{d}f=0.95
  \label{95upperlimit_Uninformative}
\end{equation}
Since $f=0$ is the mathematical singularity, in practice, we introduce a small positive constant $\epsilon$ as the lower bound for integration. 

\subsection{Updated Prior Case}\label{subsec:appendix_UpdatedPrior}
Since the data of the first observation fundamentally introduces correlation between $f$ and $\mathcal{P}$, the posterior after the first observation $p(f,\mathcal{P} | \mathrm{data_1})$ should be treated as a whole joint posterior instead of two  separate marginals. The final posterior can be written as 
\begin{eqnarray}
  p(f,\mathcal{P} | \mathrm{data}_1,\mathrm{data}_2)&&\propto\mathcal{L} (f,\mathcal{P}|\mathrm{data_1})\mathcal{L} (f,\mathcal{P}|\mathrm{data_2})\pi(f)\pi(\mathcal{P}),\nonumber\\
  &&=(1-fp)^{N_1+N_2}f^{-1}.
  \label{posteriordata12}
\end{eqnarray}
The 95\%  credible upper limit can be obtained by solving
\begin{equation}
  \int_{0}^{f_{\mathrm{upper}}}\int_{0}^{1}\frac{p(f, \mathcal{P}| \mathrm{data}_1,\mathrm{data}_2)}{p(\mathrm{data_1,data_2})}\mathrm{d}\mathcal{P}\mathrm{d}f=\int_{0}^{f_{\mathrm{upper}}}\frac{1-(1-f)^{N_1+N_2+1}}{p(\mathrm{data_1,data_2})f^2(N_1+N_2+1)}\mathrm{d}f=0.95.
  \label{95upperlimit_update}
\end{equation}
The introduction of small positive constant $\epsilon$ is also applied in solving Equation (\ref{95upperlimit_update}).
\bibliography{sample631}{}
\bibliographystyle{aasjournal}



\end{document}